# Review: Graphene oxide 2D thin films for high performance nonlinear integrated photonics

*David J. Moss*

Optical Sciences Centre, Swinburne University of Technology, Hawthorn, VIC 3122, Australia

**Abstract**

Integrated photonic devices operating via optical nonlinearities offer a powerful solution for all-optical information processing, yielding processing speeds that are well beyond that of electronic processing as well as providing the added benefits of compact footprint, high stability, high scalability, and small power consumption. The increasing demand for high-performance nonlinear integrated photonic devices has facilitated the hybrid integration of novel materials to address the limitations of existing integrated photonic platforms, such as strong nonlinear optical absorption or an inadequate optical nonlinearity. Recently, graphene oxide (GO), with its large optical nonlinearity, high flexibility in altering its properties, and facile fabrication processes, has attracted significant attention, enabling many hybrid nonlinear integrated photonic devices with improved performance and novel capabilities. This paper reviews the applications of GO to nonlinear integrated photonics. First, an overview of GO's optical properties and the fabrication technologies needed for its on-chip integration is provided. Next, the state-of-the-art GO nonlinear integrated photonic devices are reviewed, together with comparisons of the nonlinear optical performance of different integrated platforms incorporating GO. Finally, the challenges and perspectives of this field are discussed.

# 1. Introduction

By avoiding the inefficient optical-electrical-optical conversion, all-optical signal generation, amplification, and processing based on optical nonlinearities offers processing speeds that far exceed that of electrical devices [1-3], underpinning a variety of applications such as ultrafast switching [4, 5], broadband optical amplification [6, 7], optical logic gates [8, 9], all-optical wavelength conversion [10, 11], metrology [12, 13], spectroscopy [14, 15], optical cloaking [16, 17], and quantum optics [18, 19]. Compared to bulky discrete off-chip devices, photonic integrated circuits fabricated by well-established complementary metal-oxide semiconductor (CMOS) technologies provide an attractive solution to implement compact nonlinear optical devices on a chip scale, thus harvesting great dividends for integrated devices such as high stability and scalability, low power consumption, and large-scale manufacturing [20-22].

Although silicon-on-insulator (SOI) has been the dominant platform for photonic integrated circuits, its indirect bandgap is a significant handicap for optical sources, and its centrosymmetric crystal structure poses an intrinsic limitation for second-order nonlinear optical applications. Furthermore, its strong two-photon absorption (TPA) at near-infrared wavelengths limits its third-order nonlinear optical response in the telecom band [2, 23]. Other CMOS compatible platforms such as silicon nitride [7, 24, 25] and doped silica [26, 27] have a much lower TPA, although they still face the limitation of having a much smaller third-order optical nonlinearity than silicon. To address these issues, the on-chip integration of novel materials has opened up promising avenues to overcome the limitations of these existing integrated platforms. Many hybrid nonlinear integrated photonic devices incorporating polymers [28, 29], carbon nanotubes [30, 31], and two-dimensional (2D) materials [32-34] have been reported, showing significantly improved performance and offering new capabilities beyond those of conventional integrated photonic devices.

2D materials, such as graphene, black phosphorus (BP), transition metal dichalcogenides (TMDCs), hexagonal boron nitride (hBN), and graphene oxide (GO), have motivated a huge upsurge in activity since the discovery of graphene in 2004 [35]. With atomically thin and layered structures, they have exhibited many remarkable optical properties that are intrinsically different from those of conventional bulk materials [36-42]. Recently, there has been increasing interest in the nonlinear optical properties of 2D materials, which are not only fascinating in terms of laboratory research but also intriguing for potential practical and industrial applications [43-50].

Amongst the different 2D materials, GO has shown many advantages for implementing hybrid integrated photonic devices with superior nonlinear optical performance [37, 51-54]. It has been reported that GO has a large third-order optical nonlinearity ($n_2$) that is over 4 orders of magnitude higher than silicon [55, 56] as well as a linear absorption that is over 2 orders of magnitude lower than graphene in the infrared region [52, 57], both of which are very useful for third-order nonlinear optical processes. In addition, GO has a heterogenous atomic structure that exhibits non-centrosymmetry, yielding a large second-order optical nonlinearity that is absent in pristine graphene that has a centrosymmetric structure. The bandgap and defects in GO can also be engineered to facilitate diverse linear and nonlinear optical processes. These material properties of GO, together with its facile synthesis processes and high compatibility with integrated platforms [57, 58], have enabled a series of high-performance nonlinear integrated photonic devices. Here, we provide a systematic review of these devices, highlighting their applications based on a range of nonlinear optical processes (**Figure 1**) as well as a comparison of the different integrated platforms.

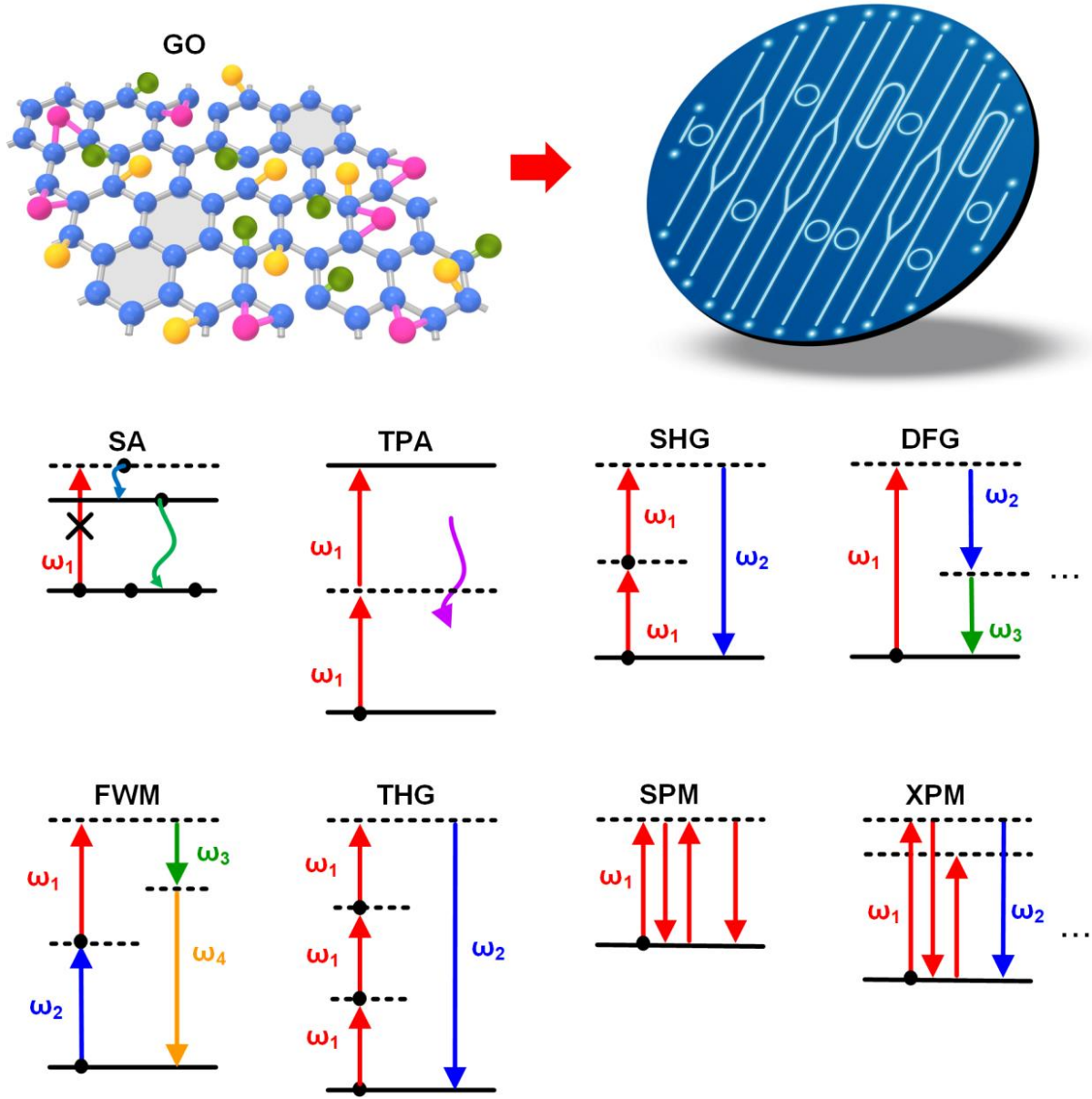

**Figure 1.** Schematic illustration of on-chip integration of GO for nonlinear optical applications. SA: saturable absorption. TPA: two-photon absorption. SHG: second-harmonic generation. DFG: difference frequency generation. FWM: four-wave mixing. THG: third-harmonic generation. SPM: self-phase modulation. XPM: cross-phase modulation.

The review paper is organized as follows. In **Section 2**, the optical properties of GO, including both the linear and nonlinear properties, are introduced, particularly in the context of integrated photonic devices. Next, the fabrication technologies for integrating GO films on chips are summarized in **Section 3**, which are classified into GO synthesis, film coating on chips, and device patterning. In **Section 4**, we review the state-of-the-art nonlinear integrated photonic devices incorporating GO. In **Section 5**, detailed comparison for the nonlinear optical performance of different integrated platforms incorporating GO is presented and discussed. In **Section 6**, the current challenges and future perspectives are discussed. Finally, the conclusions are provided in **Section 7**.

## 2. Optical properties of GO

GO, which contains various oxygen-containing functional groups (OCFGs) such as epoxide, hydroxyl, and carboxylic, all attached on a graphene-like carbon network, is one of the most common derivatives of graphene [59-62]. The heterogeneous atomic structure including both $sp^2$ carbon sites with π-states and $sp^3$-bonded carbons with σ-states makes GO exhibit a series of distinctive material properties, particularly in its 2D form. In this section, we briefly introduce GO's optical properties, including both the linear and nonlinear properties and focusing on the near-infrared telecom band (around 1550 nm). **Table 1** provides a comparison of the basic optical properties of GO with typical 2D materials such as graphene, TMDCs, and BP as well as bulk materials such as silicon (Si), silica ($SiO_2$), silicon nitride ($Si_3N_4$), and high index doped silica glass (Hydex) used for implementing integrated photonic devices. In the following, we provide detailed introduction of GO's optical properties based on **Table 1**.

**Table 1.** Comparison of basic optical properties of GO with other 2D materials and bulk materials for integrated photonic devices. LO: linear optical. NLO: nonlinear optical.

| Material | Bandgap (eV) | LO parameters [a] | | NLO parameters [a] | | Ref. |
|---|---|---|---|---|---|---|
| | | $n$ | $k$ | $n_2$ ($\times 10^{-14}$ m$^2 \cdot$W$^{-1}$) | $\beta$ ($\times 10^{-12}$ m$\cdot$W$^{-1}$) | |
| GO | 2.1 – 3.5 | 1.97 | 0.005 – 0.01 | 1.5 – 4.5 | -5.3×10$^4$ | [53, 55, 63] |
| Graphene | 0 | 2.7 | 1.37 – 2.21 | -8.0 – -10.0 | 9.0×10$^4$ | [64-67] |
| MoS$_2$ | 1.2 – 1.8 | 3.65 | 0.1 – 0.9 | 0.027 | 120 | [68-72] |
| WSe$_2$ | 1.2 – 1.7 | 3.70 | 0.04 | -0.17 | 2.1×10$^4$ | [73-76] |
| BP | 0.3 – 2.0 | 2.65 | 1.2 – 1.9 | -1.25 × 10$^{-5}$ | 0.18 | [77-83] |
| Si | 1.1 – 1.3 | 3.48 | ≈0 | 6 × 10$^{-4}$ | 5.0 | [2, 3, 84] |
| SiO$_2$ | 6.5 – 7.5 | 1.45 | ≈0 | 2.60 × 10$^{-6}$ | — | [85, 86] |
| Si$_3$N$_4$ | 4.5 – 5.3 | 1.99 | ≈0 | 2.61 × 10$^{-5}$ | — | [3, 24, 87] |
| Hydex | — | 1.6 – 1.7 | ≈0 | 1.3 × 10$^{-5}$ | — | [3, 88, 89] |

[a] Here we show the values of $n$, $k$, $n_2$, and $\beta$ around the wavelength of 1550 nm.

## 2.1 Linear optical properties

In contrast to graphene that has a bandgap of zero [90], GO has a typical bandgap between 2.1 eV – 3.6 eV [59, 91], which yields low linear light absorption in the telecom band. Although in **Table 1** the optical extinction coefficient $k$ of GO (0.005 – 0.01) is not as low as for Si, Si$_3$N$_4$, and SiO$_2$, it is nonetheless still much lower than the other 2D materials, particularly graphene where the loss is over 100 times lower. This property of GO is highly attractive for nonlinear optical applications such as self-phase modulation (SPM) and four-wave mixing (FWM) that require high power to drive the nonlinear processes. On the other hand, GO has a refractive index $n$ that is around 2 across a broad optical band from near-infrared to mid-infrared regions [53, 55, 62, 63]. This results in a low material dispersion, which is critical for implementing devices with broad operation bandwidths, e.g., broadband FWM or SPM devices based on phase matching [2, 6].

The bandgap of GO can be engineered by using different reduction methods to change the ratio of the *sp*$^2$ and *sp*$^3$ fractions [92, 93], thus yielding a variation in its material properties. **Figure 2(a)** compares the atomic structures of graphene, GO, reduced GO (rGO), and totally reduced GO (trGO). As can be seen, with the continued removal of the OCFGs, GO gradually reduces and finally converts to trGO. As compared with graphene, trGO has a similar carbon network but with more defects. The differences in the properties of trGO and graphene mainly come from these defects, which can form not only during the reduction process but also the oxidation process associated with the conversion from graphene to GO [94, 95]. **Figure 2(b)** compares the measured *n*, *k* of GO, rGO, trGO, and graphene [63, 96]. As the degree of reduction increases, both *n* and *k* of rGO increase and show a trend towards graphene, with the *n* and *k* of trGO being extremely close to those of graphene. In contrast to bulk materials that have limited tuning ranges with respect to *n* and *k* (e.g., typically on the order of $10^{-4} - 10^{-3}$ for *n* of Si [97]), GO has a very wide tuning range for both *n* (from ~2 to ~2.7) and *k* (from < 0.01 to ~2), which underpins many photonic devices that achieve excellent phase and amplitude tuning capabilities [63, 96].

Similar to graphene and TMDCs [98-100], GO films exhibit strong anisotropy in its optical absorption in a broad band from the visible to the infrared regions [57, 101]. This property is useful for implementing polarization selective devices with wide operation bandwidths.

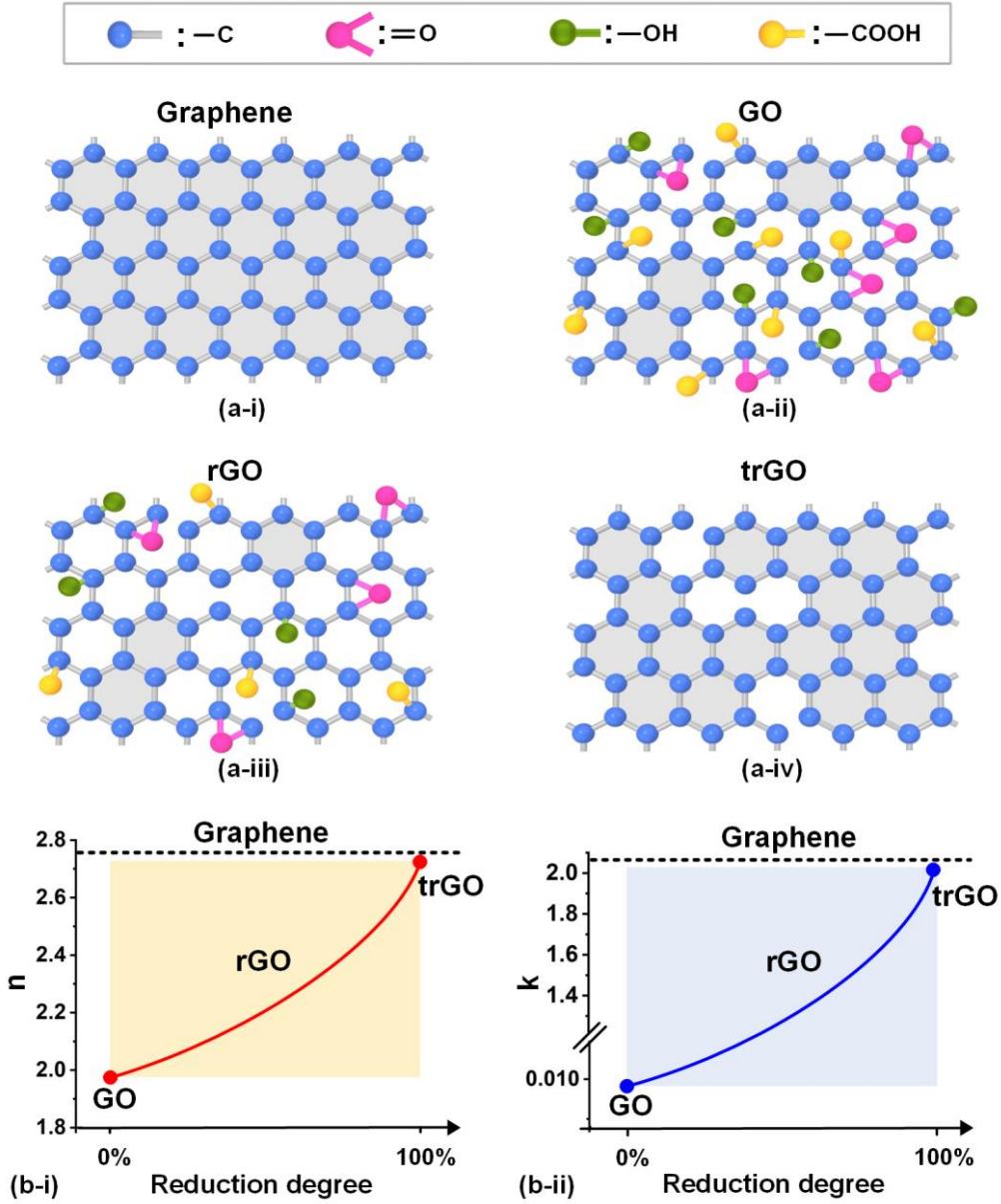

**Figure 2.** (a) Schematics of atomic structures of (i) graphene, (ii) GO, (iii) rGO, and (iv) trGO. (b) Comparison of their (i) refractive index *n* and (ii) extinction coefficient *k*.

In **Figure 3**, we compare different integrated waveguides incorporating 2D layered GO films, including Si, $Si_3N_4$, and Hydex waveguides with typical cross sections of 0.50 μm × 0.22 μm, 1.60 μm × 0.66 μm, and 2.00 μm × 1.50 μm, respectively. **Figure 3(a)** shows schematics of the hybrid waveguides. For comparison, we choose integrated waveguides with planarized top surfaces with each waveguide coated with 1 layer of GO film (~2 nm in thickness [52, 53]). Unless elsewhere specified, the bare integrated waveguides in our following discussions are the same as those in **Figure 3(a)**. **Figure 3(b)** shows the transverse electric (TE) mode profiles for the hybrid waveguides, which were calculated based on the experimentally measured *n* of 2D layered GO films at 1550 nm [63, 96]. **Figure 3(c)** compares the refractive indices of GO, Si, $Si_3N_4$, Hydex, and $SiO_2$ over the wavelength range of 1500 nm – 1600 nm. GO has a refractive index that is higher than either Hydex or $SiO_2$, but lower than $Si_3N_4$ and Si. Si has the highest refractive index amongst the three waveguide materials, which results in the tightest light confinement in the waveguide and hence the smallest waveguide geometry.

**Figure 3(d)** shows the dispersion of the bare Si, $Si_3N_4$, and Hydex waveguides without GO films. The dispersion of the GO-coated Si waveguide is also shown for comparison. All the dispersions were simulated using the refractive indices in **Figure 3(c)**. The bare Si waveguide has normal disperison, whereas the bare $Si_3N_4$ and Hydex waveguides have slight anomalous dispersion. After coating with GO films, the GO-Si hybrid waveguide has a slightly reduced normal dispersion, while the GO-$Si_3N_4$ and GO-Hydex waveguides exhibit slightly enhanced anomalous dispersion (not shown in **Figure 3(d)** since the curves for these wavegudies almost overlap with those of the uncoated waveguides), indicating that incorporating GO films could benefit phase matching for FWM or SPM in these waveguides .

**Figures 3(e)** and **(f)** show the ratios of power in GO relative to the power in the waveguide core for different numbers of GO layers *N*, respectively. The thickness of the GO film was assumed to be proportional to *N* in the simulation. For the hybrid waveguides with the same GO layer number, GO-Si waveguide has the strongest evanescent field leakage and mode overlap with the GO film , mainly a result of its smaller waveguide geometry. All the hybrid waveguides show an increased mode overlap with the GO films as *N* increases, reflecting the fact that the increase of GO film thickness can enhance the interaction between light and GO.

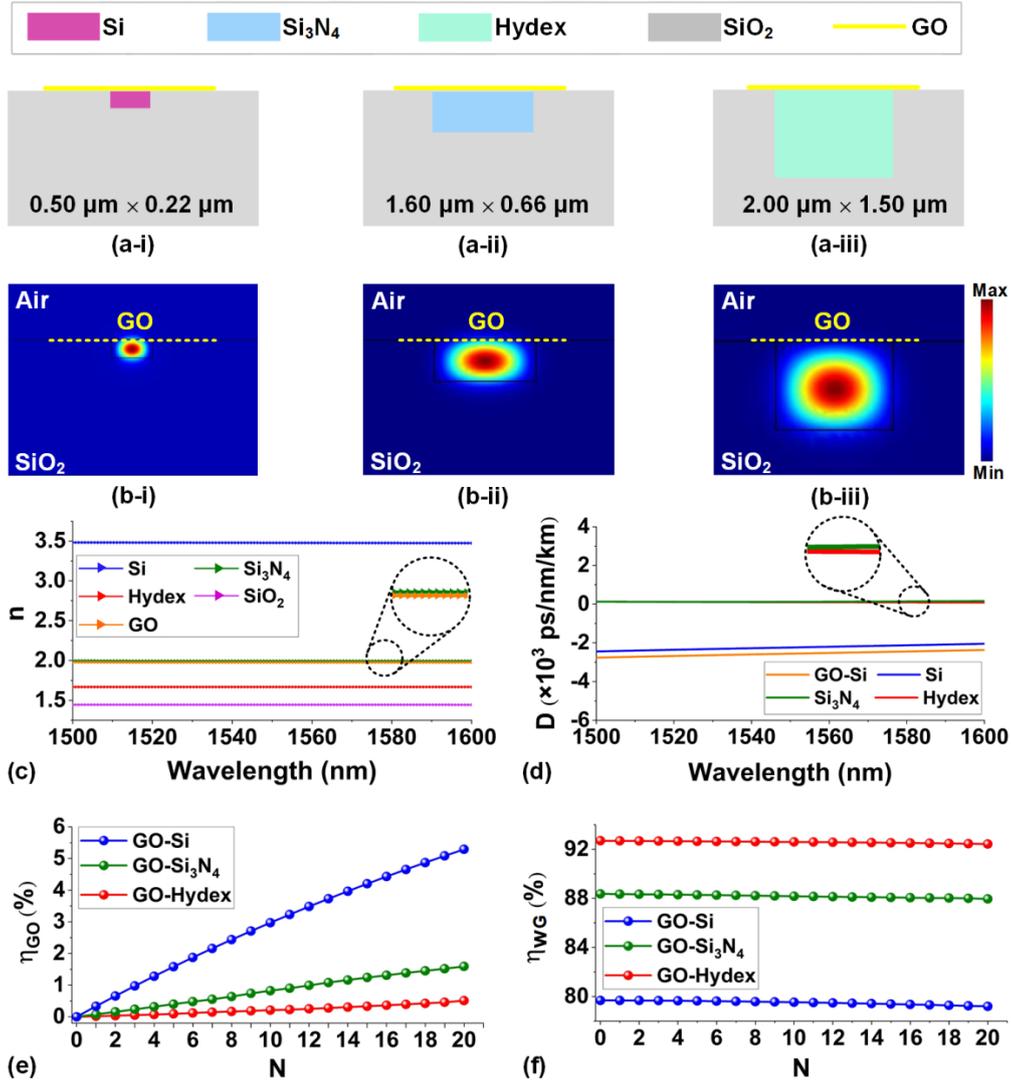

**Figure 3.** (a) Schematic illustration of cross sections for (a-i) Si (0.50 μm × 0.22 μm), (a-ii) Si$_3$N$_4$ (1.60 μm × 0.66 μm), and (a-iii) Hydex (2.00 μm × 1.50 μm) waveguides, each coated with 1 layer of GO film. (b) TE mode profiles of the hybrid waveguides in (a). (c) Comparison of refractive indices $n$ of GO, SiO$_2$, Si, Si$_3$N$_4$, and Hydex. (d) Comparison of dispersions of the three integrated waveguides without coating GO films and GO-coated Si waveguide. (e) Ratio of power in GO to that in all material regions ($\eta_{GO}$) versus layer number $N$ for different hybrid waveguides. (f) Ratio of power in the waveguide core to that in all material regions ($\eta_{WG}$) versus $N$ for different hybrid waveguides.

In **Figure 4(a)**, we compare the linear propagation loss of the hybrid waveguides versus GO layer number $N$, which was calculated based on the experimentally measured $k$ of 2D layered GO films at 1550 nm [63, 96]. For practical GO films, the value of $k$ slightly increases with $N$, which mainly results from the accumulated film imperfections induced by film unevenness, stacking of multiple layers, and localized defects [53, 57]. As can be seen, the GO-Si waveguide has a much higher propagation loss than comparable GO-Si$_3$N$_4$ and GO-Hydex waveguides, and all of these waveguides show an increased propagation loss with increasing *N*. This is similar to the results shown in **Figure 3(e)**, indicating that an enhanced GO mode overlap results in increased linear propagation loss. Mode overlap plays an important role in balancing the trade-off between enhancing the third-order optical nonlinearity while minimizing linear loss to achieve the optimized performance for the GO hybrid waveguides, which has been discussed in detail in Refs. [102, 103].

The linear propagation loss of practical GO hybrid waveguides exposed to air can change with input light power, especially at high average powers [51, 54]. Such power-dependent linear loss (PDLL) results from power-sensitive photo-thermal changes in the GO films, including a range of effects such as photo-thermal reduction, thermal dissipation, and self-heating in the GO layers [51, 54, 104]. The photo-thermal changes arising from these sources show some interesting features. First, in a certain power range where the light power is not sufficiently high to induce permanent changes in the films, the changes can recover back when the light power is turned off. Second, their time responses (typically on the order of $10^{-3}$ s [51]) are much slower than those of the ultrafast third-order nonlinear optical processes (typically on the order of $10^{-15}$ s [34, 105]). Finally, these changes are sensitive to the average light power in the GO films, and so are easily triggered by continuous-wave (CW) light with high average power. In contrast, for optical pulses with a high peak power but a low average power, the PDLL induced by these changes is not obvious [51, 54].

**Figures 4(b) – (d)** compare the excess linear propagation loss induced by the PDLL

($\Delta PL_{PDLL}$, after excluding the corresponding linear propagation loss in **Figure 4(a)**) versus average power of input CW light for the hybrid GO-Si, GO-Si$_3$N$_4$, and GO-Hydex waveguides, respectively. The $\Delta PL_{PDLL}$ increases with both the GO film thickness and the average power for all the hybrid waveguides, reflecting the fact that there are more significant photo-thermal changes in thicker GO films, and particularly at higher average powers. The GO-Si waveguide shows much higher $\Delta PL_{PDLL}$ than the GO-Si$_3$N$_4$ and GO-Hydex waveguides with the same GO layer number – a result also arising from its stronger GO mode overlap that allows for a higher power in the GO film.

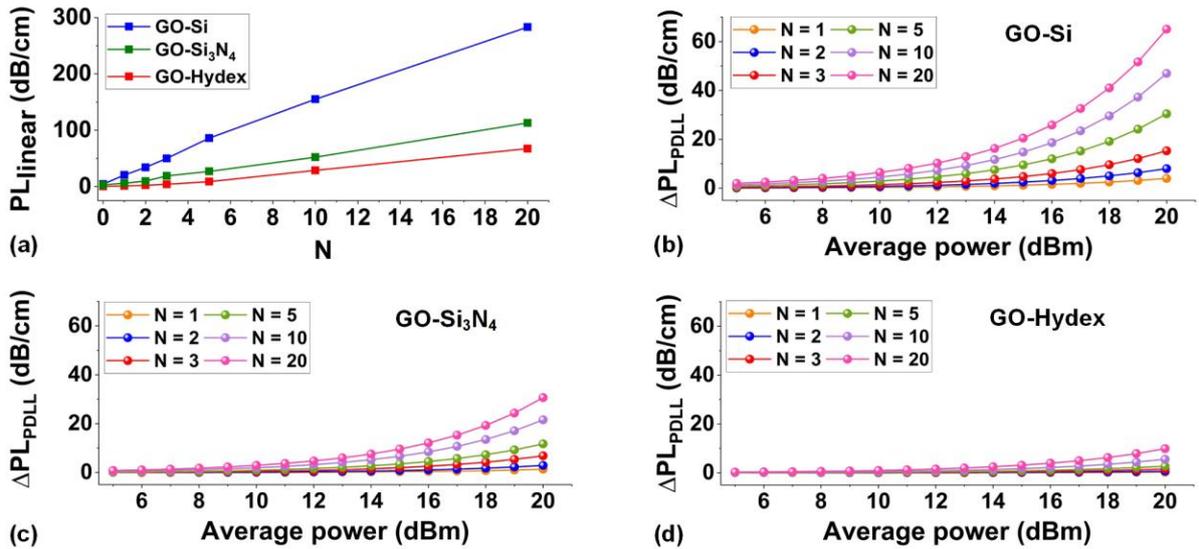

**Figure 4.** (a) Linear propagation loss ($PL_{linear}$) versus layer number $N$ for GO-Si, GO-Si$_3$N$_4$, and GO-Hydex waveguides. (b) – (d) Excess linear propagation loss induced by the PDLL ($\Delta PL_{PDLL}$) versus average power of input continuous-wave (CW) light for the hybrid Si, Si$_3$N$_4$, and Hydex waveguides coated with different numbers of GO layers, respectively.

## 2.2 Nonlinear optical properties

Upon interaction with an external optical electric field having a high intensity, on the order of interatomic fields (i.e., $10^5 – 10^8$ V/m [106]), materials can exhibit nonlinear optical responses that can be accompanied by novel phenomena such as the generation of new frequencies, or with their linear optical parameters such as $n$ and $k$ becoming field-dependent [107]. In the past decade, the superior nonlinear optical properties of 2D materials have been widely investigated and recognized [44, 45, 48, 108]. For GO, its heterogeneous structure and tunable bandgap enable distinctive nonlinear optical properties for a diverse range of nonlinear optical processes [51, 52, 54, 57].

Generally, the material's nonlinear optical properties include second-order, third-order, and higher-order responses described by the complex susceptibility tensors $\chi^{(i)}$ [2, 44], where $i$ = 2, 3, ... denote the $i^{th}$-order. In this paper, we focus on GO's third-order optical nonlinearity, highlighting the on-chip integration of GO films for both enhanced Re ($\chi^{(3)}$) and Im ($\chi^{(3)}$) processes. For the second-order optical nonlinearity, we note that large $\chi^{(2)}$ values of GO arising from its non-centrosymmetric atomic structure have been reported recently [109, 110], but its application to chip-scale devices is still in its infancy. Therefore, we provide a discussion on the future perspectives for this in Section 6.

The Re ($\chi^{(3)}$) processes (also termed parametric processes [2, 44]), represented by four-wave mixing (FWM), self- / cross-phase modulation (SPM / XPM), and third harmonic generation (THG), play an integral role in all-optical signal generation and processing with an ultrafast time response on the order of femtoseconds [10, 111, 112]. In **Table 1**, the Kerr coefficients ($n_2$) of relevant materials are also compared. The absolute value of $n_2$ for GO is about 10 times lower than that of graphene but still much higher than those of $MoS_2$, $WSe_2$, and BP. On the other hand, the $n_2$ of GO is about 4 – 5 orders of magnitude higher than Si, $Si_3N_4$, and Hydex, and so this forms the motivation for the on-chip integration of GO to implement hybrid devices for third-order nonlinear optical applications.

For some processes, such as the Kerr nonlinearity, the terms arising from Im($\chi^{(3)}$) involve nonlinear optical absorption such as two-photon absorption (TPA), saturable absorption (SA), or multi-photon absorption (MPA) [2, 44]. The relatively large bandgap of GO results in low TPA in the telecom band that is helpful for improving the efficiency for the Re ($\chi^{(3)}$) processes [2, 3]. In contrast to the TPA process where the absorption increases with light intensity, SA exhibits the opposite trend, due to the saturation of excited electrons filling the conduction band and hence preventing further transitions due to Pauli blocking [49, 113]. In **Table 1**, the negative $\beta$ for GO is induced by SA, which originates from the ground-state bleaching of the $sp^2$ domain [56, 114, 115]. The SA in GO is useful for applications such as mode-locked fiber lasers [116-118] and all-optical modulators [105, 119]. The bleaching of light absorption at high intensities is also beneficial for boosting processes arising from the Re ($\chi^{(3)}$). Compared to the photo-thermal changes mentioned in **Section 2.1**, SA is an ultrafast third-order nonlinear optical process determined by the peak input light power, and so it is more easily triggered by optical pulses with high peak powers. For high average power CW light with low peak power, the SA-induced loss change is not as observable.

As mentioned in **Section 2.1**, the bandgap of GO can be changed by using different reduction methods [92, 93]. By increasing the degree of reduction, a switch in sign for both $n_2$ and $\beta$ of GO films has been observed during the transition from GO to trGO [55, 56]. The large dynamic tunable ranges for $n_2$ and $\beta$ provide high flexibility in tailoring the performance of nonlinear integrated photonic devices incorporating GO.

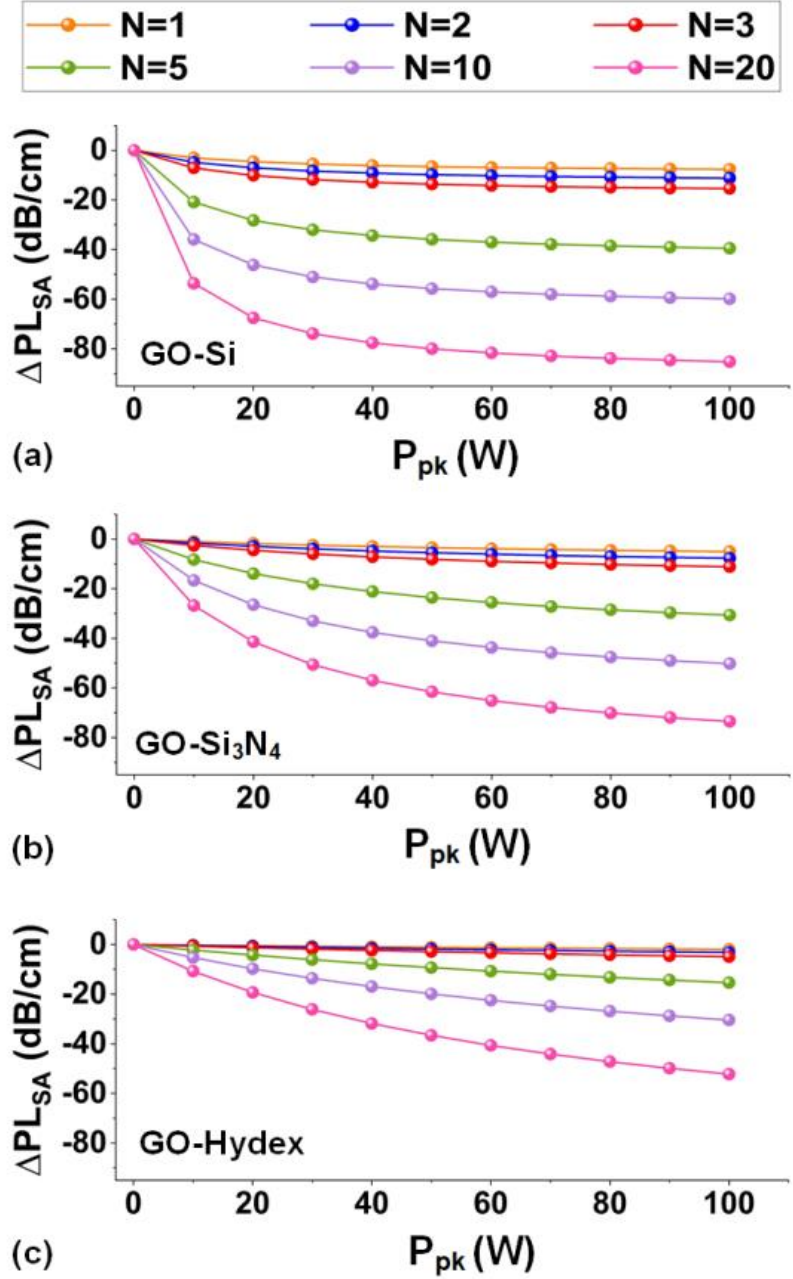

**Figure 5.** Excess propagation loss induced by the SA ($\Delta PL_{SA}$) versus peak power of input optical pulses ($P_{pk}$) for (a) Si, (b) Si$_3$N$_4$, and (c) Hydex waveguides coated with different numbers of GO layers.

**Figures 5(a) – (c)** compare the excess propagation loss induced by GO's SA ($\Delta PL_{SA}$, after excluding the corresponding linear propagation loss in **Figure 4(a)**) versus peak input power of the optical pulses ($P_{pk}$) for the hybrid GO-Si, GO-Si$_3$N$_4$, and GO-Hydex waveguides, respectively. For all the hybrid waveguides, $\Delta PL_{SA}$ becomes more significant for increasing number of layers and input peak power, reflecting more significant SA in thicker GO films and at higher peak powers. Similar to the trend seen in **Figure 4**, GO-Si waveguides with stronger mode overlap show higher $\Delta PL_{SA}$ than the GO-Si$_3$N$_4$ and GO-Hydex waveguides for the same number of GO layers.

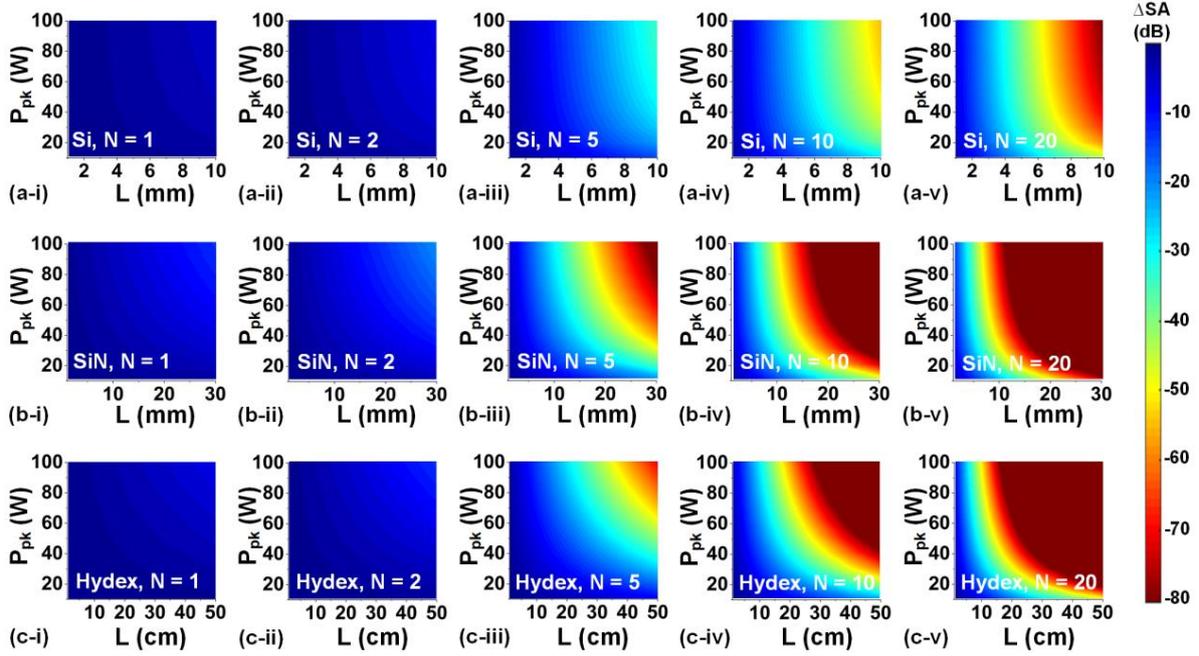

**Figure 6.** Excess insertion loss induced by the SA ($\Delta SA$) as functions of peak power of input optical pulses ($P_{pk}$) and waveguide length ($L$) for hybrid (a) Si, (b) Si$_3$N$_4$, and (c) Hydex waveguides uniformly coated with different numbers of GO layers. (i) – (v) show the results for layer number $N$ = 1, 2, 5, 10, and 20, respectively.

**Figures 6(a) − (c)** compare the overall excess insertion loss induced by SA ($\Delta SA$), after excluding the corresponding linear insertion loss, as functions of $P_{pk}$ and waveguide length $L$ for the uniformly coated GO-Si, GO-Si$_3$N$_4$, and GO-Hydex waveguides, respectively. In each figure, the results for 5 different GO layer numbers are provided. To highlight the difference, different ranges for the waveguide length were chosen in **Figures 6(a) – (c)**. It is seen that $\Delta SA$ increases with both GO layer number and input peak power, which is consistent with **Figure 5**. In addition, $\Delta SA$ also increases with waveguide length, reflecting a more significant SA-induced insertion loss difference for longer waveguides.

## 3. On-chip integration of GO films

The distinctive material properties of GO have motivated its on-chip integration for implementing functional hybrid integrated devices [57, 120-122]. The facile solution-based synthesis process of GO and its high compatibility with integrated device fabrication offer competitive advantages for industrial manufacturing beyond the laboratory, which has thus far been a challenge for the majority of 2D materials. In this section, we review the fabrication techniques for integrating GO films on chips, which are divided into GO synthesis, film coating on chips, and device patterning.

### 3.1 GO synthesis

Material synthesis is the first step before integrating GO films onto chips. In contrast to graphene that has very low solubility in water, GO can be dispersed in aqueous and polar solvents, thus allowing for solution-based material synthesis. The Brodie method [123] and the Hummer method [124] are the two basic GO synthesis approaches, both of which have long histories and have been modified on the basis of the initially proposed methods [58, 125]. **Figures 7(a)** and **(b)** show schematic illustrations of these two methods. For the Brodie method, graphite is treated with fuming nitric acid ($HNO_3$) and potassium chlorate ($KClO_3$) in order to attach the OCFGs (**Figure 7(a)**), whereas for the Hummer method, the oxidation of graphite is achieved via treatment with potassium permanganate ($KMnO_4$) and sulfuric acid ($H_2SO_4$) (**Figure 7(b)**). Compared to the Brodie method, the Hummer method is more facile and shows better compatibility with CMOS fabrication technologies. **Figures 7(c)** and **(d)** show another two GO synthesis approaches that are well-known for the GO community – the Staudenmaier method and the Hofmann method. Both of these are modifications of the Brodie method, with slight changes in the procedure intended to produce highly oxidized GO [126, 127]. The former uses a mixture of concentrated fuming $HNO_3$ and $H_2SO_4$ followed by adding $KClO_3$, whereas the latter uses concentrated $HNO_3$ in combination with concentrated $H_2SO_4$ and $KClO_3$. Some modified Hummer methods have also been proposed [58, 128], where the amount of $KMnO_4$ and $H_2SO_4$ were engineered to improve the oxidation efficiency and hence the oxidation degree.

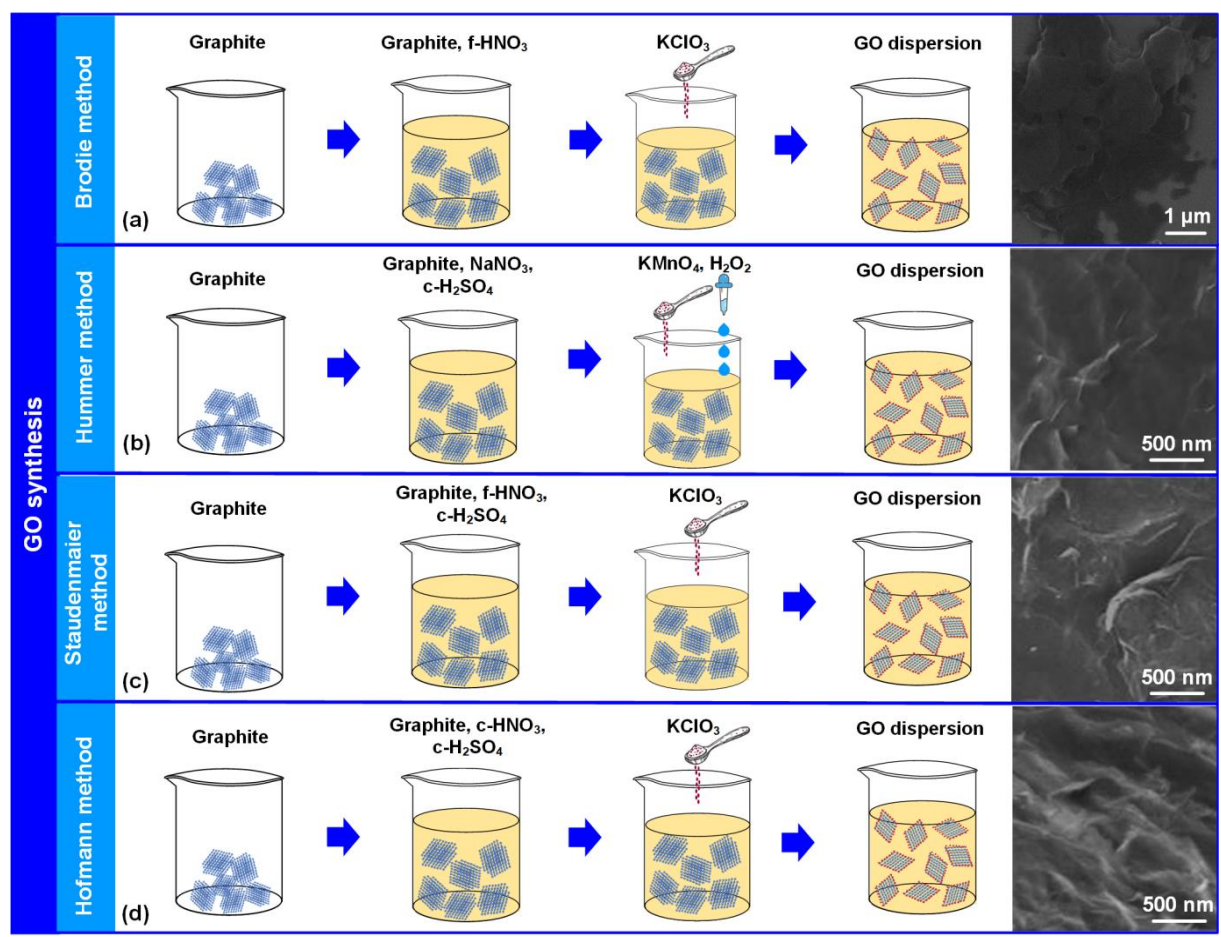

**Figure 7.** Schematic illustration of typical GO synthesis methods: (a) the Brodie method, (b) the Hummer method, (c) the Staudenmaier method, and (d) the Hofmann method. c-: concentrated. f-: fuming. In (a) – (d), the figure in the right side of each row shows an image for as-fabricated samples. The sample image in (a) is reproduced with permission.[125] Copyright 2013 Elsevier Ltd. The sample images in (b) – (d) are reproduced with permission.[129] Copyright 2014 Wiley-VCH.

The above methods can produce a large volume of exfoliated GO sheets with a high concentration of OCFGs, which are easily disintegrated into smaller flakes. The lateral size (typically varying from several tens of nanometers to several tens of microns) and thickness (typically on the order of nanometers) of the GO flakes can be controlled by varying the mixing or sonication parameters. GO films consisting of large-size (>10 μm) flakes show better performance in terms of electrical / thermal conductivity as well as mechanical / sieving capability, whereas GO films made from small-size flakes are advantageous in achieving conformal coating on substrates with complex structures, particularly for integrated devices having feature sizes on the micron or nanometer scale.

**3.2 Film coating on chips**

The second step is to coat GO films onto integrated chips. In contrast to sophisticated film transfer processes used for the on-chip integration of graphene and TMDCs, the coating of GO films can be realized using solution-based methods without any transfer processes. **Figure 8** shows schematic illustrations of two typical GO film coating strategies – solution dropping and self-assembly. Both of these are compatible with the Brodie and the Hummer methods and are suited to large-scale fabrication, but they also show differences, particularly with respect to film uniformity and thickness.

Solution dropping methods, mainly including drop casting [130] and spin or spray coating [131, 132], are simple and rapid to directly coat GO films in large areas. The main steps in these methods include solution preparation, solution dropping, and drying (**Figure 8(a)**). The relatively low film uniformity and large film thicknesses are the main limitations for these methods, which make it challenging to achieve film conformal coating of integrated waveguides. The typical film unevenness that they produce is > 10 nm, and the typical film thicknesses are > 100 nm [101, 133].

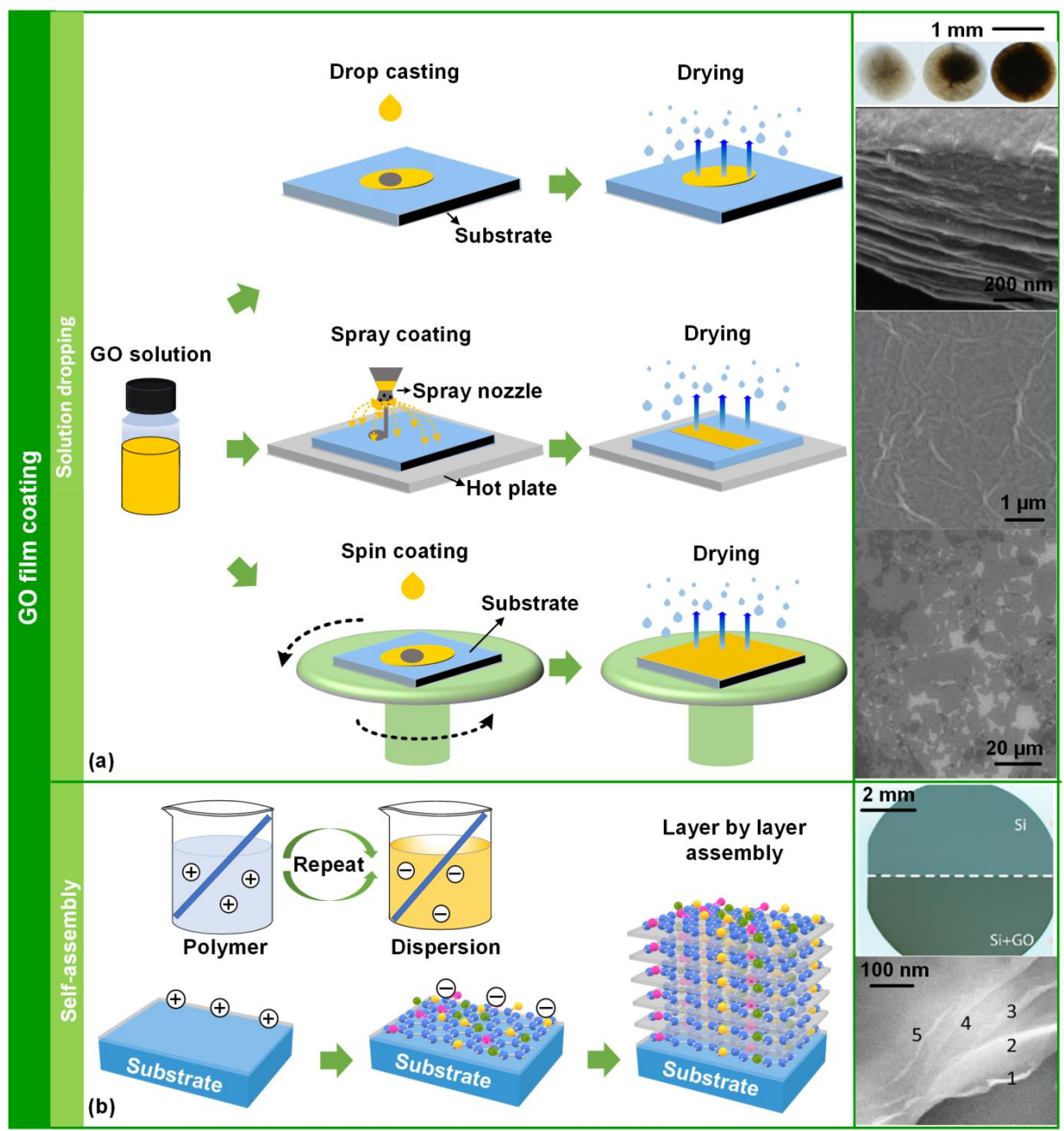

**Figure 8.** Schematic illustration of typical methods for GO film coating: (a) solution dropping and (b) self-assembly. In (a) and (b), the figures in the right side of each row show images for as-fabricated samples. The sample images in (a) are reproduced with permission.[101] Copyright 2014, OSA Publishing [134] Copyright 2018, Elsevier Ltd, and [132] Copyright 2010, American Chemical Society. The sample images in (b) are reproduced with permission.[63] Copyright 2019, American Chemical Society.

In contrast to solution dropping methods, self-assembly methods can achieve both high film uniformity (< 1 nm [63]) and low film thickness (down to the thickness of 2D monolayers [57]). **Figure 8(b)** shows the process flow for self-assembly. First, a GO solution composed of negatively charged 2D GO nanoflakes synthesized via the Brodie or the Hummer methods is prepared. Second, the target integrated chip with a negatively charged surface is immersed in a solution with positively charged aqueous polymers to obtain a polymer-coated integrated chip with a positively charged surface. Finally, the polymer-coated integrated chip is immersed in the prepared GO solution, where a GO monolayer is formed onto the top surface through electrostatic forces. By repeating the above steps, layer-by-layer coating of GO films on integrated chips can be realized, with high scalability and accurate control of the layer number or the film thickness. The strong electrostatic forces also enable conformal film coating of complex structures (e.g., wire waveguides and gratings) with high film uniformity. In addition, unlike film transfer approaches where the coating areas are limited by the lateral size of the exfoliated 2D films [135, 136], the film coating area for the self-assembled methods is limited only by the size of the substrate and the solution container, which makes them excel at large-area film coating. By using plasma oxidation, the removal of GO films coated from integrated devices can be easily achieved, allowing for the recycling of the integrated chips and recoating of new GO films.

**3.2 Device patterning**

Device patterning is critical for engineering functionalities of advanced integrated devices. In **Figure 9**, we summarize the typical methods used to pattern GO films, including inkjet printing, laser writing, lithography followed by lift-off, pre-patterning, and nanoimprinting. All of these methods have strong potential for industrial manufacturing, and each of them has advantages for specific applications. In **Table 2**, we compare the different GO film patterning methods.

Inkjet printing is a simple and rapid GO film patterning method that can simultaneously achieve film coating and patterning. It is compatible with solution dropping coating methods, and is usually employed to fabricate patterns over large areas, with relatively low resolution on the order of microns [137, 138]. **Figure 9(a)** shows the process flow, where specialized ink solutions need to be prepared before printing. The printing processes involve forming a jet of single droplets, drop casting, and droplet drying, similar to solution dropping methods, with the pattern shape and position normally controlled via programs.

**Table 2. Comparison of various GO films patterning methods.**

| Technique | Resolution | Speed | Prefabrication [a] | Resist | Etching free | Refs. |
|---|---|---|---|---|---|---|
| Inkjet printing | > 1 μm | Fast | No | No | Yes | [137, 138] |
| Laser writing | > 300 nm | Moderate | No | No | Yes | [96, 139] |
| Lithography & lift-off | — [b] | Fast | Yes | Yes | No | [54, 57] |
| Pre-patterning | > 200 nm | Fast | Yes | No | Yes | [140] |
| Nanoimprinting | 10 – 100 nm | Moderate | Yes | Yes | No | [141] |

[a] This includes prefabricated masks, moulds, and substrates.
[b] The patterning resolution, typically ranging from several nanometers to several hundreds of nanometers, depends on the light wavelength used for lithography.

Laser writing is a one-step, noncontact, and mask-free film patterning method that has been widely used for patterning polymers [142, 143], metal surfaces [144, 145], and 2D materials [146, 147]. **Figure 9(b)** illustrates the process flow for patterning GO films using laser writing. The laser source can consist either of CW or pulsed lasers, with an objective lens used to focus the laser beam. Laser writing involves complex processes such as GO reduction, thermal melting / sublimation, and structural reorganization, ultimately resulting in localized thinning or ablation of GO films depending on the laser power. Laser writing can be used to pattern both thick films deposited by solution dropping and thin films coated via self-assembly, The patterning resolution is mainly determined by the spot size of the focused laser beam, which typically ranges from several microns to hundreds of nanometers [148].

Lithography followed by lift-off is another widely used GO film patterning method [54, 57]. Unlike laser wrting that peforms patterning and etching simultanously, for lithography the patterns are first formed on photoresist using well-developed techniques in the integrated circuit industry such as photolithography and electron beam lithography. It is then transferred to GO films deposited on the photoresist via lift off processes that are common for fabricating integrated metal electrodes (**Figure 9(c)**). Compared to GO films coated via solution dropping, films formed by self-assembly show a better lift-off outcome owing to their strong adhesion to the substrates enabled by the electrostatic forces. The patterning resolution is determined by both the lithography resolution and the film property. For visible, ultraviolet, deep ultraviolet (DUV) photolithography, the patterning resolution is mainly limited by the lithography resolution (typically > 300 nm), which is much larger than the sizes of the exfoliated GO nanoflakes (typically ~50 nm). For electron beam lithography with a higher patterning resolution (typically ~100 nm), the influence of the GO film thickness and flake size becomes more prominent, especially when the minimum feature size is < 150 nm [57].

Direct coating of GO films onto pre-patterned structures, or pre-patterning, is a simple method that can realize large-area GO film patterning. It relies on pre-fabrication to pattern the target substrates and conformal coating of GO films (**Figure 9(d)**), thus being suitable for the self-assembled GO films [140]. Pre-patterning is usually used for mass producing repetitive patterns. Similar to lithography followed by lift off, the patterning resolution is mainly limited by the minimum gap width of the pre-patterned structure when it is > 300 nm, and the GO film thickness and flake size when it is < 150 nm.

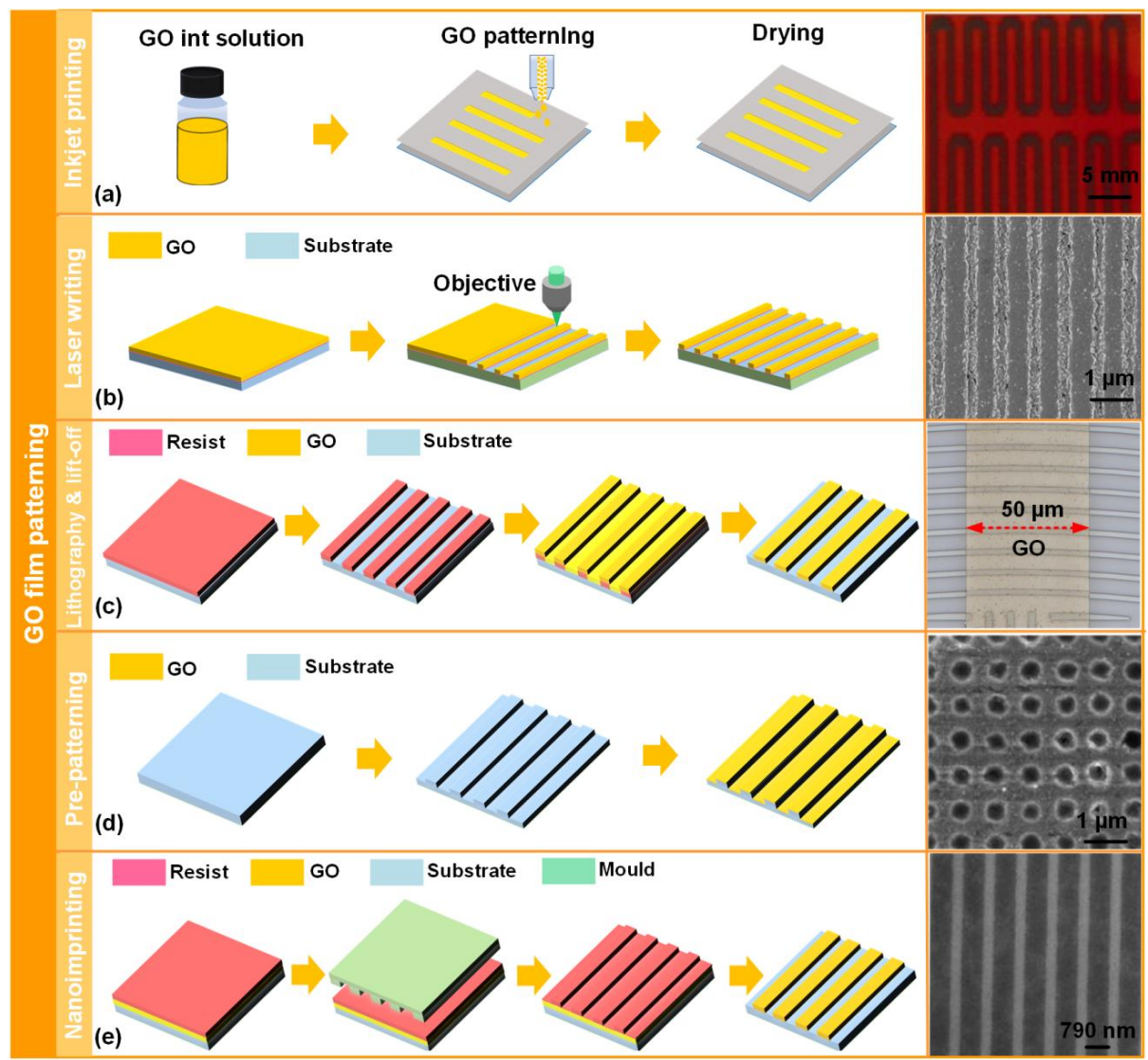

**Figure 9.** Schematic illustration of typical methods for patterning GO films: (a) inkjet printing, (b) laser writing, (c) lithography & lift-off, (d) pre-patterning, and (e) nanoimprinting. In (a) – (e), the figure in the right side of each row shows an image for as-fabricated samples. The sample images in (a) – (e) are reproduced with permission.[149] Copyright 2011 Springer Nature., [120] Copyright 2019, Springer Nature, [54] Copyright 2020, Wiley-VCH, [140] Copyright 2020, Springer Nature, and [150] Copyright 2011 American Vacuum Society, respectively.

Nanoimprinting is a film patterning method that can achieve a very high patterning resolution (e.g., down to ~10 nm [151]). Similar to lithography followed by lift off, it also needs to pattern photoresist before transferring the patterns onto the GO films. Instead of using photolithography or electron beam lithography, prefabricated imprint molds are employed to pattern the photoresist (**Figure 9(e)**). To fabricate different patterns, different molds are required, and so this approach is mainly used for fabricating relatively simple and repetitive patterns [141].

Finally, it is worth mentioning that the fabrication techniques used to incorporate GO films in **Figures 7 – 9** are not limited to nonlinear integrated photonic devices. Rather, they are universal and can be used to fabricate other integrated photonic devices such as polarizers [57, 152], lenses [153, 154], and sensors [155, 156], and also integrated electronic devices such as field-effect transistors [155, 157], supercapacitors [158, 159], and solar cells [160, 161]. For nonlinear integrated photonic devices, self-assembly methods are more widely used than solution dropping methods, mainly due to the high film uniformity and low film thickness they can achieve, which result in low film loss that is desirable for boosting nonlinear optical processes such as FWM and SPM.

## 4. Enhanced nonlinear optics in GO hybrid integrated devices

The large optical nonlinearity and low loss of GO, along with its facile fabrication processes for large-scale and highly precise on-chip integration, have enabled many hybrid integrated devices with superior nonlinear optical performance [51-54, 110, 162]. In this section, we summarize the state-of-the-art nonlinear integrated photonic devices incorporating GO.

As a typical third-order process, FWM has been widely exploited for all-optical signal generation, amplification, and processing [6, 24, 163-165]. Enhanced FWM in GO hybrid integrated devices was first demonstrated using Hydex waveguides [53], where FWM measurements were performed for 1.5-cm-long waveguides uniformly coated with 1 – 5 layers of GO, and a maximum net conversion efficiency (CE) enhancement of 6.9 dB was achieved for the device with 2 layers of GO (**Figure 10(a)**).

Enhanced FWM in Hydex micro-ring resonators (MRRs) with patterned GO films was subsequently demonstrated [54]. Benefitting from the resonant enhancement in the MRRs, a maximum CE enhancement of 10.3 dB was achieved for a MRR with a patterned film including 50 GO layers (**Figure 10(b)**). Based on the FWM measurements, the change in $n_2$ of GO films as a function of the layer number and light power was also analyzed, showing interesting trends in evolving from 2D materials to bulk-like behavior. Following the experimental demonstration, detailed theoretical analyses and optimization were performed in Ref. [166], showing that CE enhancement up to 18.6 dB can be achieved by optimizing the GO coating length and coupling strength of the MRR.

Enhanced FWM in GO-$Si_3N_4$ waveguides has also been demonstrated [51], where FWM measurements were carried out for GO-coated planarized $Si_3N_4$ waveguides having different GO film lengths and thicknesses, achieving a maximum CE improvement of 9.1 dB for a device with a 1.5-mm-long patterned film including 5 GO layers (**Figure 10(c)**). The patterned device also showed a broadened conversion bandwidth compared to the uncoated and uniformly coated devices. A detailed analysis of the influence of the GO film parameters and the $Si_3N_4$ waveguide geometry was provided in Ref. [102], showing that the CE enhancement can be further increased to 20.7 dB and the conversion bandwidth can be improved by up to 4.4 times.

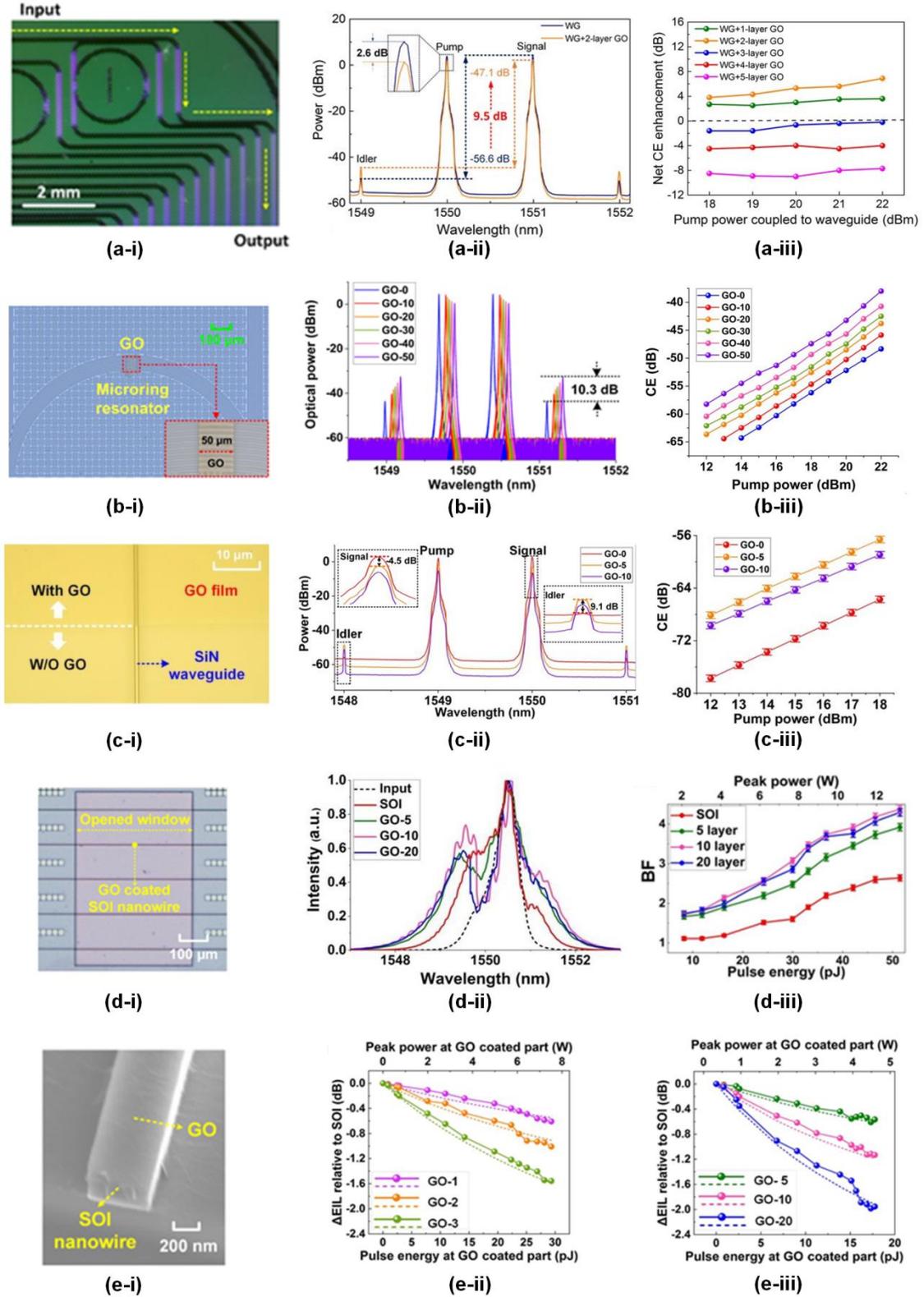

**Figure 10.** (a) Enhanced four-wave mixing (FWM) in GO-coated Hydex waveguides. (b) Enhanced FWM in GO-coated Hydex micro-ring resonators (MRRs). (c) Enhanced FWM in GO-coated $Si_3N_4$ waveguides. (d) Enhanced self-phase-modulation (SPM) in GO-coated Si waveguides. (e) Strong saturable absorption (SA) in GO-coated Si waveguides. In (a) – (d), (i) shows a microscope image for the fabricated device incorporating GO, (ii) shows measured FWM or SPM spectra for bare and GO-coated devices, and (iii) shows FWM conversion efficiency (CE)

versus input pump power or spectral broadening factors (BFs) versus input pulse energy for the devices in (ii). In (e), (i) shows a scanning electron microscopy (SEM) image of a Si waveguide conformally coated with 1 layer of GO. (ii) and (iii) show the power-dependent excess insertion loss relative to the bare Si waveguide (ΔEIL) versus pulse energy. (a) is reproduced with permission.[53] Copyright 2018, AIP Publishing. (b) is reproduced with permission.[54] Copyright 2020, Wiley-VCH. (c) is reproduced with permission.[51] Copyright 2020, Wiley-VCH. (d) and (e) are reproduced with permission.[52] Copyright 2020, American Chemical Society.

SPM is another fundamental third-order process that has wide applications in wideband optical sources, pulse compression, frequency metrology, and optical coherence tomography [167, 168]. Enhanced SPM in GO-Si waveguides has been reported [52], where SPM measurements were performed for Si wire waveguides conformally coated with GO films having different lengths and thicknesses. Significant spectral broadening of picosecond optical pulses after passing these waveguides was observed, showing a maximum broadening factor (BF) of 4.34 for a device with 10 GO layers (**Figure 10(d)**). By coating GO films, the effective nonlinear figure of merit (FOM) of the hybrid waveguide was improved by up to 20 times compared to the bare Si waveguide. According to theoretical calculations based on the experimental results [103], a maximum BF of 27.8 can be achieved by optimizing the GO film parameters and Si waveguide geometry. In addition to enhanced SPM, strong SA in the GO-coated Si waveguide was also observed, as evidenced by a decrease in the measured excess insertion loss relative to the bare Si waveguide for an increased pulse energy (**Figure 10(e)**). It was also observed that the hybrid waveguides with thicker GO films showed a more prominent SA, although at the expense of higher linear loss.

## 5. Comparison of different integrated platforms incorporating GO

As reviewed in Section 4, enhanced nonlinear optical responses have been achieved for integrated Si, $Si_3N_4$, and Hydex devices incorporating GO. In this section, we provide a detailed comparison of the nonlinear optical performance of these integrated platforms. We compare

FWM using CW light and SPM-induced spectral broadening using optical pulses in GO-coated Si, Si$_3$N$_4$, and Hydex waveguides. We used the material parameters obtained from experimental measurements [51-53] to calculate the FWM and SPM performance parameters based on the theory in Refs. [169-172], and accounted for the variation in loss arising from photo-thermal changes and SA in the GO films. The comparison of the nonlinear figure-of-merits for different GO hybrid waveguides are also provided.

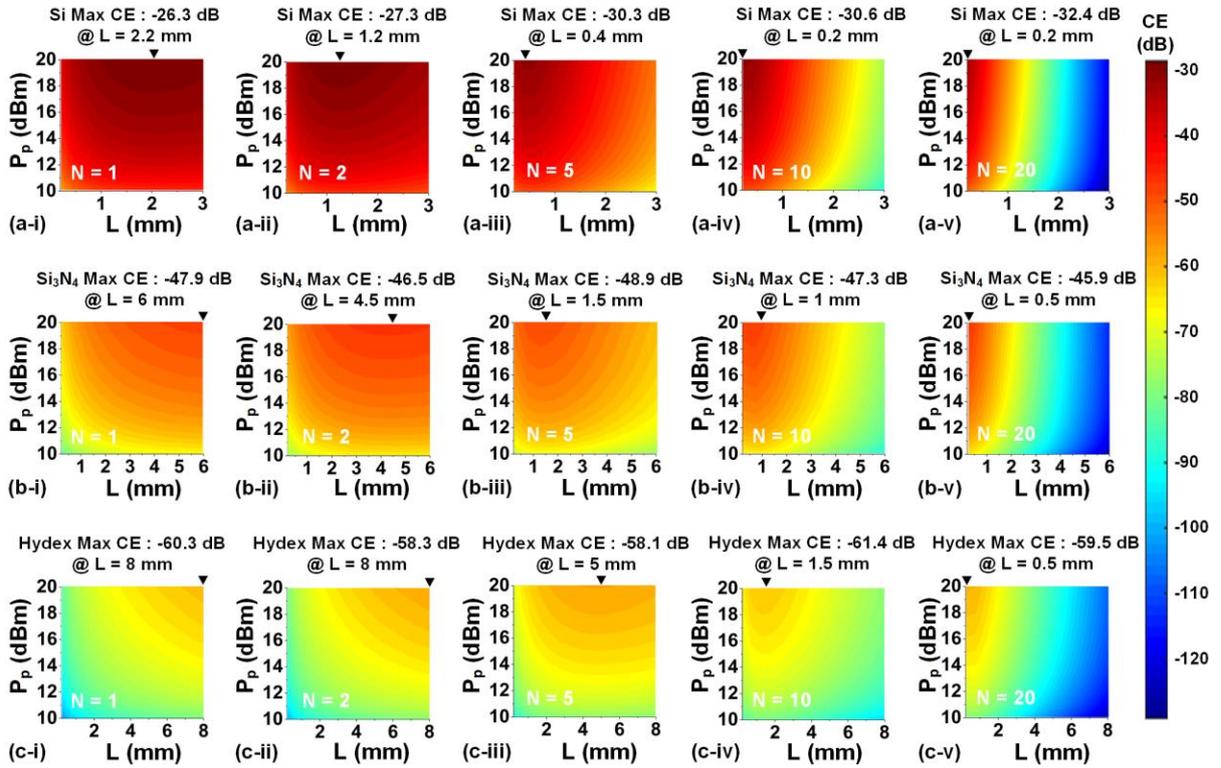

**Figure 11.** FWM CE versus waveguide length ($L$) and pump power ($P_p$) for hybrid (a) Si, (b) Si$_3$N$_4$, and (c) Hydex waveguides uniformly coated with different numbers of GO layers. (i) – (v) show the results for $N$ = 1, 2, 5, 10, and 20, respectively.

**Figure 11** shows the FWM CE as a function of waveguide length $L$ and pump power $P_p$ for hybrid Si, Si$_3$N$_4$, and Hydex waveguides uniformly coated with GO films. Similar to **Figure 6**, we show the results for 5 different numbers of GO layers (i.e., $N$ = 1, 2, 5, 10, 20). For each of the hybrid waveguides, the CE increases with $P_p$, while as a function of $L$, it first increases and then decreases, reaching a maximum value at an intermediate waveguide length. As the layer number $N$ increases, the $L$ corresponding to the maximum CE becomes smaller. These trends reflect the trade-off between third-order nonlinearity improvement and propagation loss increase for the hybrid waveguides, with the former dominating for relatively small $N$ and $L$, and the latter becoming more obvious as $N$ and $L$ increase. For the waveguides with the same $N$, the CE of the GO-Si waveguide is much higher than the GO-Si$_3$N$_4$ and GO-Hydex waveguides, although its waveguide length is shorter. This is mainly due to the larger third-order optical nonlinearity of Si as compared with Si$_3$N$_4$ and Hydex.

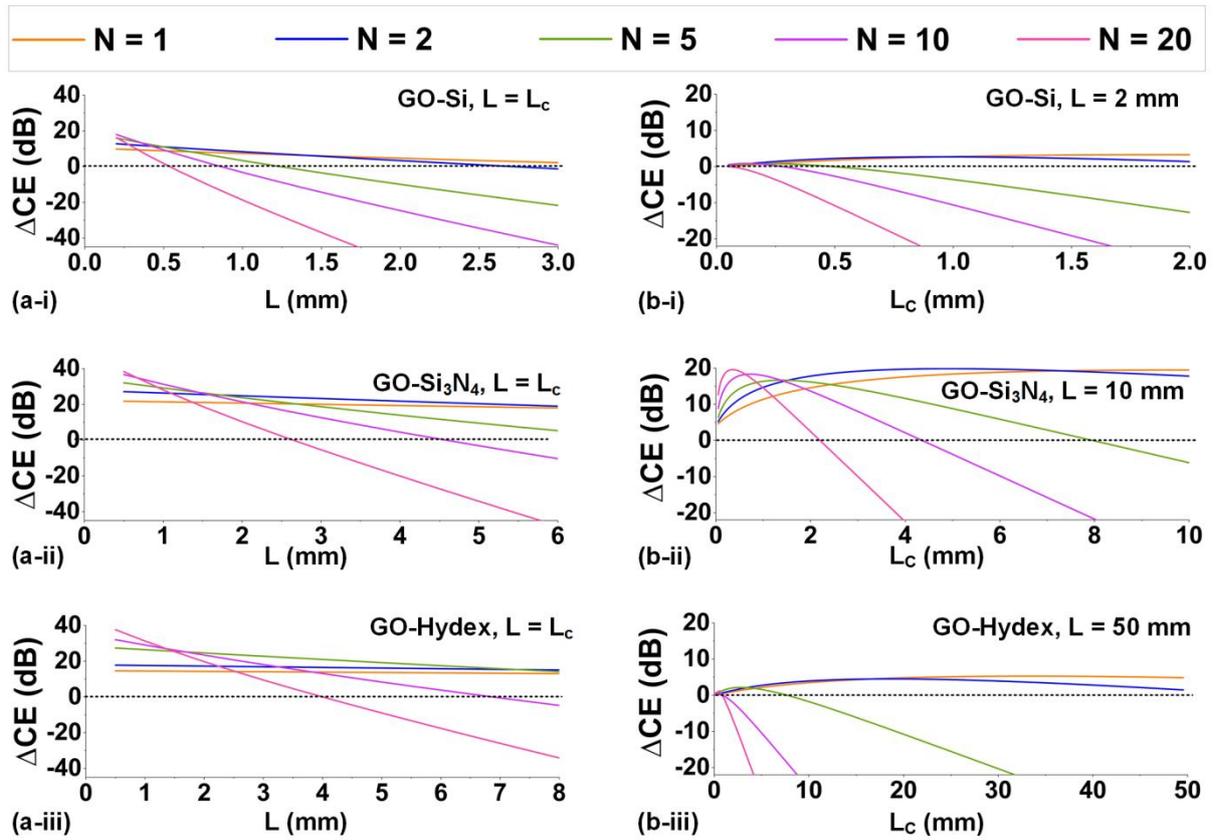

**Figure 12.** (a) FWM CE enhancement versus waveguide length ($L$) for hybrid Si, Si$_3$N$_4$, and Hydex waveguides uniformly coated with different numbers of GO layers. (b) FWM CE enhancement versus GO coating length ($L_c$) for hybrid Si, Si$_3$N$_4$, and Hydex waveguides patterned with different numbers of GO layers. The lengths of the uncoated Si, Si$_3$N$_4$, and Hydex waveguides are $L$ = 2 mm, $L$ = 10 mm, and $L$ = 50 mm, respectively. The patterned GO films are assumed to be coated from the start of the waveguides.

**Figure 12** compares the CE enhancement ($\Delta CE$) of the hybrid waveguides relative to the uncoated waveguides. In **Figure 12(a)**, we show the results for the waveguides uniformly coated with GO films. For all of these hybrid waveguides, the CE enhancement decreases with waveguide length $L$, reflecting the fact that a shorter length yields better CE enhancement. For the waveguides coated with thicker GO films, although the initial CE enhancement (at very small $L$) is higher, it decreases more rapidly with $L$, thus resulting in a decreased range of $L$ with positive CE enhancement. **Figure 12(b)** presents the corresponding results for the waveguides with patterned GO films, where the length of the uncoated waveguide is fixed at $L$, and the GO film coating length $L_c$ varies from 0 to $L$. Similar to the relation between CE and $L$ in **Figure 11**, the CE enhancement reaches a maximum for an intermediate $L_c$, and the $L_c$ corresponding to the highest $\Delta CE$ decreases with GO layer number $N$. This also results from the trade-off between the third-order nonlinearity and loss. The CE enhancement of GO-Si waveguides is lower than the GO-Si$_3$N$_4$ and GO-Hydex waveguides, in contrast to a higher CE achieved for the GO-Si waveguides in **Figure 11**. This reflects an interesting trade-off between achieving high relative CE enhancement versus high overall CE in these GO hybrid integrated waveguides.

**Figure 13** shows SPM-induced spectral evolution of optical pulses traveling along the hybrid Si, Si$_3$N$_4$, and Hydex waveguides uniformly coated with GO films. For comparison, we show the BFs at an intensity attenuation of -20 dB (i.e., $BF_{-20dB}$) for different waveguides, together with the corresponding propagation lengths ($L_p$). For each of the hybrid waveguides, $BF_{-20dB}$ first increases and then decreases with GO layer number $N$, achieving a maximum spectral broadening at an intermediate film thickness. As $N$ increases, $L_p$ decreases and the optical pulses vanish at shorter propagation lengths, reflecting that the loss increase becomes dominant for the waveguides with longer lengths or thicker GO films. Similar to **Figure 11**, the large third-order optical nonlinearity of Si result in the GO-Si waveguides showing a much more significant spectral broadening than comparable GO-Si$_3$N$_4$ and GO-Hydex waveguides even for shorter lengths. Unlike the symmetric spectral evolution in the GO-Si$_3$N$_4$ and GO-Hydex waveguides, the spectral evolution in the GO-Si waveguides exhibits a slight asymmetry due to free-carrier effects in Si.

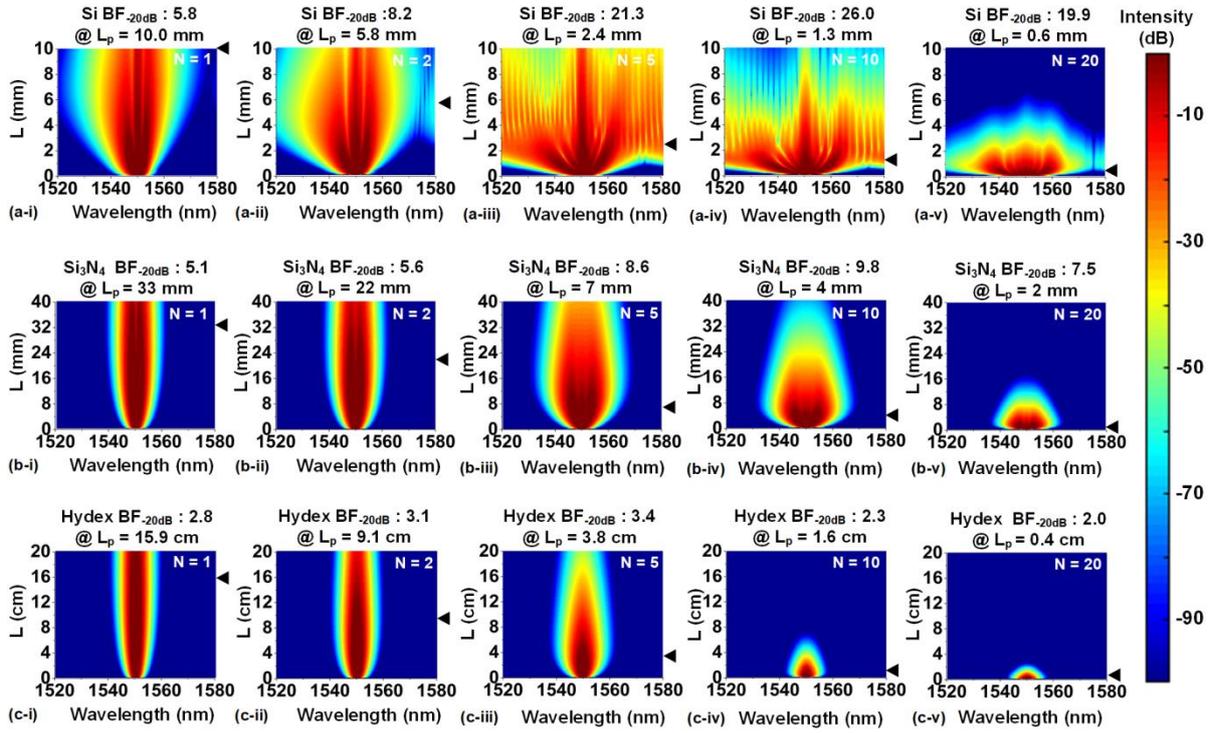

**Figure 13.** SPM-induced pulse spectral evolution for hybrid (a) Si, (b) Si$_3$N$_4$, and (c) Hydex waveguides uniformly coated with different numbers of GO layers. (i) – (v) show the results for $N$ = 1, 2, 5, 10, and 20, respectively. The full width at half maximum (FWHM) and peak power of the input optical pulses are 3.9 ps and 30 W, respectively.

**Figure 14(a)** compares the relative BF (*rBF*) versus waveguide length (*L*) for hybrid Si, Si$_3$N$_4$, and Hydex waveguides uniformly coated with GO films, where the *rBF* is defined as the ratio of the BF of the hybrid waveguide to that of the uncoated waveguide. Note that the BF here corresponds to the value at the waveguide output, which is different from *BF*$_{-20dB}$ in **Figure 13**. For the GO-Si and GO-Si$_3$N$_4$ waveguides with thicker GO films ($N \geq 5$), the maximum *rBF* is achieved for an intermediate *L*, whereas for these waveguides with thinner GO films and all the GO-Hydex waveguides, the *rBF* monotonically increases with *L*. This is consistent with the trade-off between the third-order nonlinearity and loss in **Figure 13**. **Figure 14(b)** compares the *rBF* versus GO coating length ($L_c$) for the waveguides with patterned GO films, where the length of the uncoated waveguide is fixed at *L*, with $L_c$ varying from 0 to *L*. Similar to the trend in **Figure 14(a)**, the maximum *rBF* is also achieved for an intermediate $L_c$ for the GO-Si and GO-Si$_3$N$_4$ waveguides when $N \geq 5$. In contrast to the higher BF achieved for the GO-Si waveguides in **Figure 13**, their *rBF* is lower than the GO-Si$_3$N$_4$ waveguides, which is similar to the trade-off between achieving a high relative CE enhancement and a high overall CE in **Figures 11** and **12**.

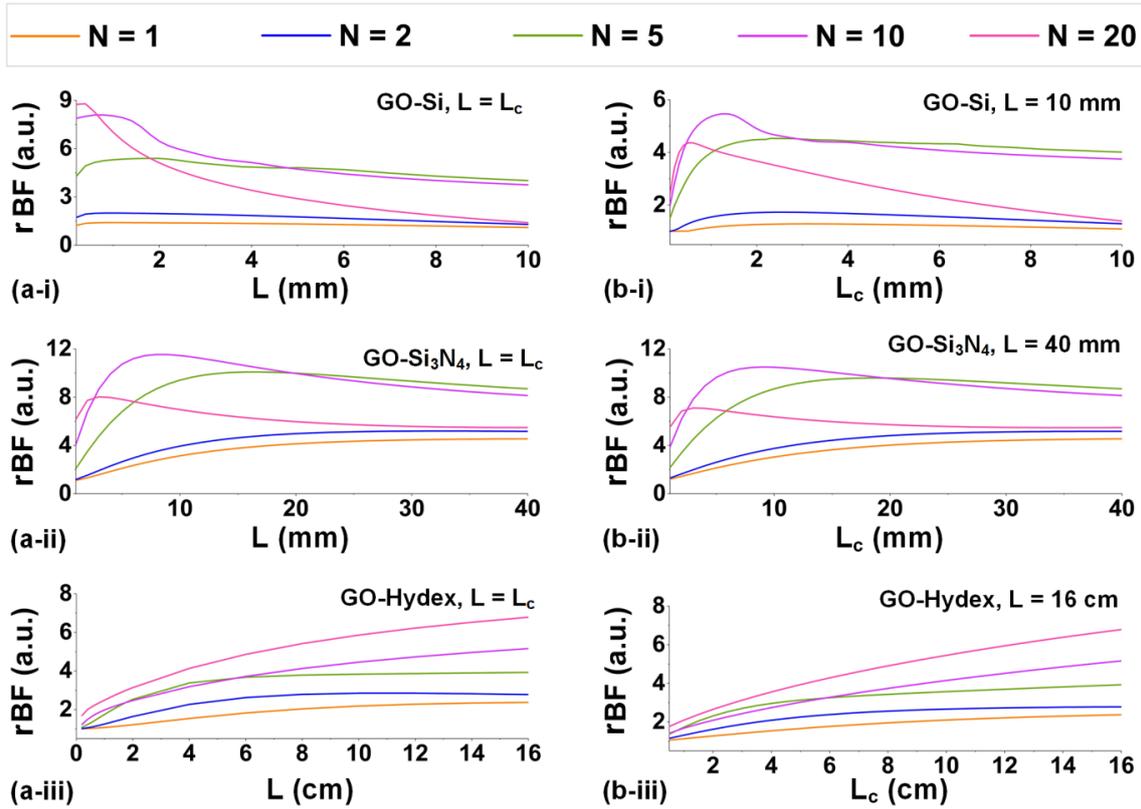

**Figure 14.** (a) SPM-induced relative BF (*rBF*) versus waveguide length (*L*) for hybrid Si, Si$_3$N$_4$, and Hydex waveguides uniformly coated with different numbers of GO layers. (b) *rBF* versus GO coating length ($L_c$) for hybrid Si, Si$_3$N$_4$, and Hydex waveguides patterned with different numbers of GO layers. The lengths of the uncoated Si, Si$_3$N$_4$, and Hydex waveguides are $L$ = 10 mm, $L$ = 40 mm, and $L$ = 16 cm, respectively. The patterned GO films are assumed to be coated from the start of the waveguides. In (a) and (b), the parameters of the input optical pulses are the same as those in Figure 13.

In **Table 3** and **Figure 15**, we quantitively compare the nonlinear performance of GO-Si, GO-Si$_3$N$_4$, and GO-Hydex waveguides, together with corresponding results for the bare integrated waveguides. We calculated two figure-of-merits, i.e., $FOM_1$ and $FOM_2$, which are widely used for comparing nonlinear optical performance. $FOM_1$ is defined in terms of nonlinear absorption [2, 3], and the corresponding results are provided in **Table 3**. It increases with GO layer number $N$, and the $FOM_1$ of the GO-Si waveguide is lower than comparable GO-Si$_3$N$_4$ and GO-Hydex waveguides. The former results from the increase in the third-order optical nonlinearity and the latter is due to the strong TPA of Si. $FOM_2$ is defined based on the trade-off between third-order optical nonlinearity and linear loss [173]. It is a function of waveguide length $L$ given by $FOM_2 = \gamma \times L_{eff}$, where $L_{eff} = [1 - exp(-\alpha_L \times L)]/\alpha_L$ is the effective interaction length, with $\gamma$ and $\alpha_L$ denoting the waveguide nonlinear parameter and the linear loss attenuation coefficient, respectively. **Figures 15(a)** and **(b)** shows $L_{eff}$ and $FOM_2$ versus $L$ for hybrid Si, Si$_3$N$_4$, and Hydex waveguides uniformly coated with 1 and 10 GO layers, respectively. Different ranges of $L$ are chosen and the results for the bare waveguides (i.e., $N = 0$) are also shown. For all of these hybrid waveguides, $FOM_2$ first rapidly increases with $L$ and then increases more gradually as $L$ becomes larger. For a small $L$, the $FOM_2$ of the hybrid waveguide is lower than that of comparable uncoated waveguide, whereas when $L$ becomes large enough, the $FOM_2$ of the uncoated waveguide gradually approaches and even surpasses that of the hybrid waveguides. This reflects that the negative influence induced by increased linear loss becomes more dominant as $L$ increases, which are consistent with the results in **Figures 11 − 14**. In contrast to $FOM_1$, the $FOM_2$ of GO-Si waveguides is higher than comparable GO-Si$_3$N$_4$ and GO-Hydex waveguides, mainly due to the large third-order optical nonlinearity of Si and its strong GO mode overlap.

**Table 3.** Comparison of nonlinear optical performance of different integrated waveguides incorporating GO. FOM: figure of merit.

| Integrated waveguide | GO layer number [a] | Waveguide dimension (μm) | $\gamma$ ($W^{-1}m^{-1}$) [b] | $\alpha_L$ (1/m) [c] | PL (dB/cm) [d] | $FOM_1$ (a. u.) [e] |
|---|---|---|---|---|---|---|
|  | $N = 0$ |  | 288.00 | 99.00 | 4.30 | 0.74 |
| Si | $N = 1$ | 0.50 × 0.22 | 668.01 | 571.03 | 24.80 | 2.52 |
|  | $N = 10$ |  | 2905.92 | 3640.30 | 158.11 | 8.90 |
|  | $N = 0$ |  | 1.51 | 69.08 | 3.00 |  |
| $Si_3N_4$ | $N = 1$ | 1.60 × 0.66 | 13.14 | 139.30 | 6.05 | >> 1 |
|  | $N = 10$ |  | 167.14 | 1474.56 | 64.04 |  |
|  | $N = 0$ |  | 0.28 | 5.53 | 0.24 |  |
| Hydex | $N = 1$ | 2.00 × 1.50 | 0.61 | 29.01 | 1.26 | >> 1 |
|  | $N = 10$ |  | 7.65 | 294.26 | 12.78 |  |

[a] $N = 0$ corresponds to the results for the uncoated Si, $Si_3N_4$, and Hydex waveguides, whereas $N = 1$ and 10 correspond to the results for the hybrid waveguides with 1 and 10 layers of GO, respectively.

[b] $\gamma$ is the nonlinear parameter. For the hybrid waveguides, $\gamma$'s are the effective values calculated based on Refs. [51, 53].

[c] $\alpha_L$ is the linear loss attenuation coefficient. According to Refs. [51-53], $\alpha_L$'s of the uncoated Si, $Si_3N_4$, and Hydex waveguides (i.e., $N = 0$) are assumed to be 99.0 $m^{-1}$, 69.1 $m^{-1}$, and 5.5 $m^{-1}$, respectively.

[d] PL is the linear propagation loss calculated by PL (dB/cm) = 10 × $\log_{10}[exp(\alpha_L \times 0.01)]$.

[e] The definition of $FOM_1 = n_2 / (\lambda \beta_{TPA})$ is the same as those in Refs. [2, 3], with $n_2$ and $\beta_{TPA}$ denoting the effective Kerr coefficient and TPA coefficient of the waveguides, respectively, and $\lambda$ the light wavelength. The values for the $Si_3N_4$, and Hydex waveguides are >> 1 due to negligible TPA observed in these waveguides.

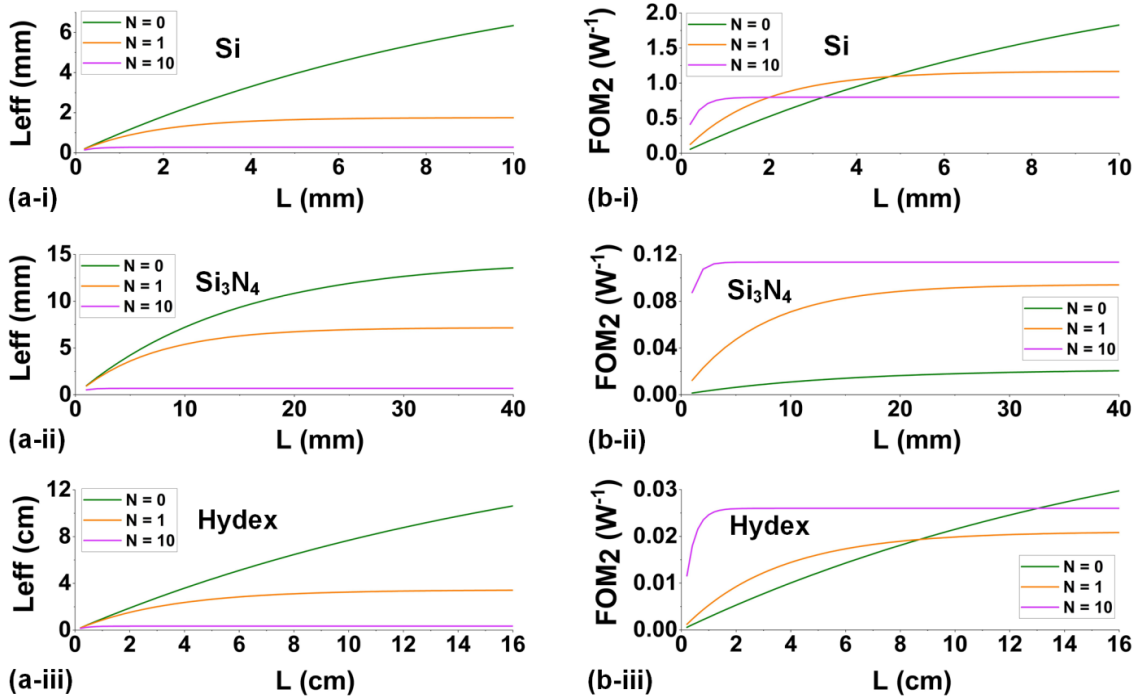

**Figure 15.** (a) Effective interaction length ($L_{eff}$) versus waveguide length ($L$) for hybrid Si, Si$_3$N$_4$, and Hydex waveguides uniformly coated with 1 and 10 layers of GO. (b) $FOM_2$ versus waveguide length ($L$) for hybrid Si, Si$_3$N$_4$, and Hydex waveguides uniformly coated with 1 and 10 layers of GO. In (a) and (b), the corresponding results for uncoated waveguides ($N$ = 0) are also shown for comparison.

## 6. Challenges and perspectives

As discussed above, GO has distinct nonlinear optical properties with excellent compatibility with different integrated platforms, yielding many high-performance hybrid nonlinear integrated photonic devices. Despite the current successes, however, there is still room for future improvement. In this section, we discuss the challenges and perspectives for fully exploiting the significant potential of GO for nonlinear integrated photonics.

Most of the state-of-the-art GO nonlinear integrated photonic devices incorporate GO films with little modification or optimization of their properties. The reality, however, as discussed in **Section 2**, is that GO's properties can be significantly changed by manipulating the OCFGs. This offers a high degree of flexibility in engineering its properties for different nonlinear optical processes. For example, a large optical bandgap of GO could benefit the FWM, SPM, and XPM processes by reducing the linear loss and nonlinear loss such as TPA. Whereas for SA, a small optical bandgap is often needed to enhance the light absorption and improve the modulation depth.

As shown in **Figure 16**, the methods for tuning GO's material properties can be classified into two categories – reduction and doping. The reduction based methods can involve thermal [174], laser [96, 175], chemical [176, 177], or microwave based reduction [178]. Amongst them, thermal and laser reduction are simple and rapid, but usually suffer from limitations in terms of residual OCFGs and generated defects. Whereas chemical and microwave reduction show better capability in completely removing the OCFGs and well preserving the carbon network without introducing many defects [174, 175]. By using wet chemistry and microwave reduction methods [178, 179], synthesizing high-quality rGO with properties extremely close to those of graphene has been realized. The synthesis of graphene-like materials via GO reduction can exploit the advantages offered by GO fabrication processes including a high production yield and a high CMOS compatibility, providing a viable solution for the mass production of graphene based devices.

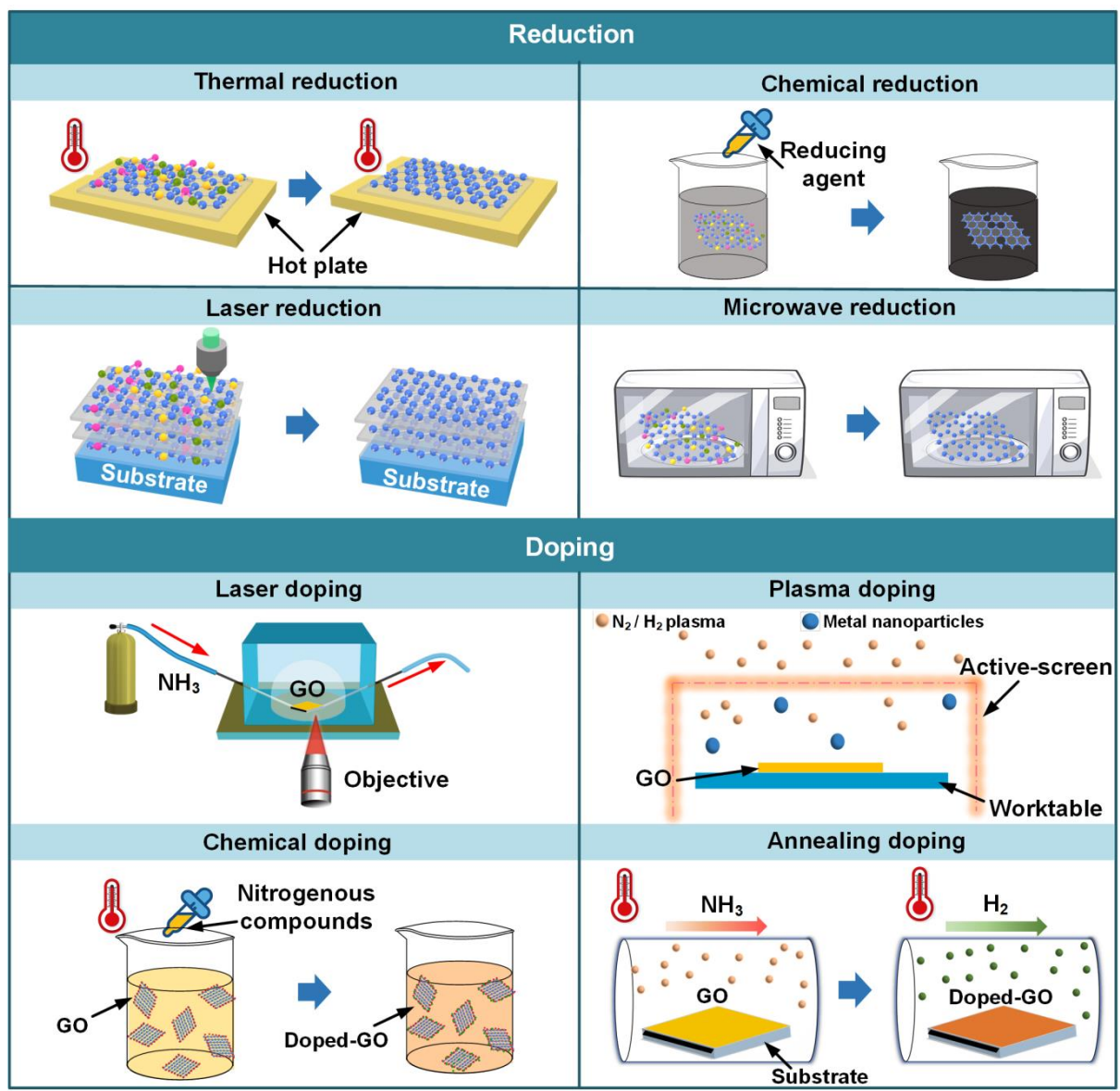

**Figure 16.** Schematic illustration of typical GO reduction and doping methods.

In contrast to the removal of OCFGs that occurs during the reduction of GO, doping methods introduce foreign atoms such as nitrogen, boron, and sulfur into the chemical structure of GO, thus enabling new material properties. The doping methods mainly consist of laser [180], chemical [181], plasma [182], and annealing based doping [183]. For laser doping, the doped area can be well controlled and patterned with a focused laser beam, but is often challenging for patterning large areas in a short time. In contrast, plasma, chemical, and annealing doping methods have shown strong ability to achieve high-efficient GO doping over large areas at the expense of a low patterning accuracy [181-183].

Although the linear propagation loss of the state-of-the-art integrated waveguides incorporating GO films is already over 100 times lower than comparable devices incorporating graphene, there is still significant room to reduce the loss even further. In principle, GO with a bandgap > 2 eV has negligible linear absorption below its bandgap, e.g., at near-infrared wavelengths (with a photon energy of ~0.8 eV at 1550 nm). The light absorption of practical GO films is mainly caused by defects as well as scattering loss stemming from imperfect layer contact and film unevenness [53, 57]. The loss from these sources can be reduced further by modifying the GO synthesis and film coating processes, e.g., by using GO solutions with improved purity and optimized flake sizes. Reducing the linear loss of the GO films will not only enhance the performance of state-of-the-art GO FWM and SPM devices, but also facilitate many new nonlinear optical applications such as supercontinuum generation (SCG) [167, 184] and optical micro-comb generation [88, 185].

The current research on GO nonlinear integrated photonic devices mainly focuses on their large third-order optical nonlinearity. However, in addition to this, a surprising second-order optical nonlinearity of GO has been reported [109, 110], which will underpin future research on GO devices for many second-order optical nonlinear processes such as second-harmonic generation (SHG), sum / difference frequency generation (SFG / DFG), the Pockels effect, and optical rectification,. Unlike the third-order optical nonlinearity that exists for all materials, the second-order optical nonlinearity can only occur in non-centrosymmetric materials or at the surface of centrosymmetric materials where the inversion symmetry is broken [44, 186, 187]. In contrast to graphene that has a centrosymmetric atomic structure, GO has a highly heterogenous atomic structure that yields a large second-order optical nonlinearity that can be tuned by changing the atomic structure of GO via reduction or doping methods. This, along with its high compatibility with integrated platforms, will enable functional second-order nonlinear integrated photonic devices with many applications, such as ultrafast signal processing and generation based on the SHG [44], tunable terahertz plasmon generation based on the DFG [188], high-speed electro-optic modulators based on the Pockels effect [189], and broadband photo detectors based on the optical rectification [190].

The use of GO films as saturable absorbers in mode-locked fiber lasers has already been demonstrated [4, 44], and the SA in integrated waveguides incorporating with GO has also been observed [52]. Although GO has a large optical bandgap with relatively low SA as compared with graphene, its SA capability can be improved by engineering the defect states in GO or by reducing it to obtain a graphene-like material. In the near future, integrated photonic devices incorporating GO or rGO with strong SA capability are expected to open new horizons for implementing on-chip mode-locked lasers [191], broadband all-optical modulators [192], pulse compression systems [193], and photonic neural networks [194].

In **Table 4**, we compare integrated photonic devices incorporating different kinds of 2D materials for nonlinear optical processes. Relative to integrated waveguides incorporating other 2D materials, GO hybrid waveguides have lower linear propagation loss. Further, the highly precise control of the film size and thickness, along with the capability for conformal coating on complex structures, offers unique advantages in optimizing the performance of GO hybrid integrated devices. Given the high flexibility in changing the material properties of GO, the potential for performance optimization is even larger. In future research on GO nonlinear integrated photonic devices, some unexplored third-order optical nonlinear processes such as THG and XPM, various second-order nonlinear optical processes, and the comparison between the performance of nonlinear integrated photonic devices incorporating rGO and graphene will potentially be very hot topics, and similar work on integrated photonic devices incorporating graphene and TMDCs in the **Table 4** could provide useful guidance for the design of experiments and for performance comparison.

The overall nonlinear optical performance of GO hybrid integrated photonic devices depends on many factors related to GO's material properties, including not only nonlinear properties such as second-order or third-order optical nonlinearity and nonlinear light absorption, but also linear properties such as linear loss and dispersion. It is complicated by effects such as changes in GO's material properties with light power and film thickness. In the meantime, these extraordinary properties of layered GO films also yield a lot of new capabilities that cannot be achieved with conventional integrated devices made from only bulk materials, which allow more degrees of freedom to engineer the device performance and functionality.

**Table 4. Comparison of integrated photonic devices incorporating different 2D materials for nonlinear optical (NLO) applications. G: graphene.**

| 2D Material | Device [a] | PL [b] (dB/mm) | 2D film thickness | $n_2$ of 2D material ($\times 10^{-14}$ m$^2$ W$^{-1}$) | NLO effect [c] | NLO performance [d] | Ref. |
|---|---|---|---|---|---|---|---|
| GO | Hydex WG | 0.1 | 4 nm 2 layers | 1.5 | FWM | CE improvement up to 6.9 dB | [53] |
| GO | Hydex MRR | — | 2−100 nm 1−50 layers | 1.2 – 2.7 | FWM | CE improvement up to 10.3 dB | [54] |
| GO | Si$_3$N$_4$ WG | 0.6 | 2−20 nm 1−10 layers | 1.3 – 1.4 | FWM | CE improvement up to 9.1 dB | [51] |
| GO | Si WG | 2.5 | 2−40 nm 1−20 layers | 1.2 – 1.4 | SPM & SA | Max BF of 4.34 Obvious SA | [52] |
| G | Si PhC cavity | — | monolayer | 10 | FWM | CE improvement more than 20 dB | [195] |
| G | Si PhC WG | 50 | monolayer | 10 | FWM | CE improvement up to 8 dB | [196] |
| G | Si MRR | — | monolayer | 15 | FWM | CE improvement up to 6.8 dB | [197] |
| G | Si$_3$N$_4$ WG | 50 | monolayer | — | FWM | $\gamma$ improvement up to about 1000 folds | [198] |
| G | Si WG | 30 | monolayer | 100 | FWM | CE improvement up to 4.8 dB | [173] |
| G | Si WG | 67−111 | monolayer | 10 | FWM | $\gamma$ improvement up to 10 folds | [199] |
| G | Si WG | 52 | — | — | SPM | Pulse compression from 80 to 15.7 fs | [200] |
| G | Si WG | 200 | monolayer | — | SPM | $\gamma$ improvement from 150 to 510 W$^{-1}$m$^{-1}$ | [171] |
| G | Si$_3$N$_4$ WG | 12.9 | monolayer | — | SA | Strong SA for varied pulse width from 0.2 to 2 ps | [172] |
| G | Si slot WG | 940 | monolayer | — | SA | MD improvement up to 7 folds [e] | [201] |
| G | Si WG | 49 | monolayer | — | SA | MD up to 22.7% | [202] |
| G | Si$_3$N$_4$ WG | 100 | 2 layers | — | DFG | Generating THz plasmons from 4.7 to 9.4 THz | [188] |
| MoS$_2$ | Si WG | 5.5 | 3 nm | 0.03 | FWM | CE improvement up to 4 dB | [68] |
| MoS$_2$ | Si WG | — | 10 nm | 0.01 | SPM | Extracted $n_2$ of MoS$_2$ about 100 times that of Si | [203] |
| MoS$_2$ | Si WG | — | monolayer | — | SHG | SHG enhancement up to 5 folds [f] | [204] |
| WS$_2$ [g] | Si$_3$N$_4$ WG | 2.5 | 5.6 nm | 0.20 – 0.22 | SPM | $\gamma$ improvement from 1.75 to 650 W$^{-1}$ m$^{-1}$ | [205] |
| WS$_2$ | Si$_3$N$_4$ WG | — | monolayer | — | SHG | Obvious SHG in contrast to negligible for the bare WG | [206] |
| GaS | Si$_3$N$_4$ MRR | — | 103 nm | $1.2 \times 10^{-4}$ | XPM | Extracted $n_2$ of GaS about 10 times that of Si$_3$N$_4$ | [207] |

[a] WG: waveguide. MRR: microring resonator. PhC: photonic crystal.

[b] PL: the linear propagation loss of the hybrid waveguides with the lowest 2D film thickness.

[c] DFG: difference frequency generation. SHG: second-harmonic generation. XPM: cross-phase modulation.

<sup>d)</sup> CE: conversion efficiency. BF: broadening factor. γ: nonlinear parameter. MD: modulation depth. The improvements in CE and $\gamma$ are relative to the uncoated waveguides.

<sup>e)</sup> The MD improvement is relative to a graphene-on-Si$_3$N$_4$ rib waveguide.

<sup>f)</sup> The SHG enhancement is relative to the free-space SHG from the 2D material.

<sup>g)</sup> This work was demonstrated for wavelength around 800 nm. Except for this work, all the other works were demonstrated for wavelengths around 1550 nm.

As discussed in **Section 2**, there are photo-thermal changes in the GO films that result in the PDLL in practical GO films. In contrast to the reversible photo-thermal changes at low light powers, the loss increase induced by the photo-thermal changes can become permanent at high powers that exceed certain thresholds. The permanent loss increase of GO limits its use for high-power nonlinear optical applications. Recently, it was found that by using an electrochemical method to modify the degree of oxidation of GO [208], it can retain a high third-order optical nonlinearity with significantly improved (> 100 times) material stability under high-power laser illumination. In future work, the on-chip integration of this modified GO is promising to yield GO hybrid integrated photonic devices with superior power handling capability.

For practical GO films under light irradiation, different nonlinear optical processes coexist [209, 210] and the interplay between these can result in complex behaviors. For example, the loss induced by TPA can deteriorate the FWM and SPM performance, whereas the reduced loss arising from SA could have a positive effect. In addition, different nonlinear optical processes may have different excitation conditions. For example, TPA occurs when the photon energy of incident light is larger than half of the material's bandgap, whereas SA can be efficiently excited when the single photon energy of incident light is just above the bandgap energy. In practical applications, the different nonlinear optical processes in GO need to be appropriately managed and balanced depending on the specific nonlinear optical application and the wavelength region of interest. For instance, the bandgap of GO can be engineered via reduction or doping to meet the requirements of specific nonlinear optical applications in specific wavelength regions.

Assembling different 2D materials to construct van der Waals heterostructures has ushered in many significant breakthroughs in recent years [211, 212]. Due to the ease of fabrication and high flexibility for changing its properties, GO offers vast possibilities for implementing heterostructures based on different materials. Currently, some heterostructures including GO or rGO have been investigated, e.g., polymer / GO heterostructure [213], titanium carbide / rGO heterostructure [214], and vanadium pentoxide / rGO heterostructure [215]. However, the optical nonlinearity of GO or rGO heterostructures, particularly in the form of integrated devices, are yet to be investigated, hinting at more significant breakthroughs to come.

Phase matching is a prerequisite for achieving efficient nonlinear processes such as FWM, SPM, XPM, and THG. For GO-coated integrated waveguides, their waveguide dispersion can be engineered by reducing or patterning GO films to alleviate the phase mismatch. This would improve the FWM bandwidth and the SPM spectral broadening, and pave the way for broadband frequency comb generation [216, 217] and SCG [184, 218]. In materials with a positive Kerr coefficient $n_2$ (e.g., Si, $Si_3N_4$ and Hydex glass), phase matching occurs for anomalous dispersion. This requires growing thick films to achieve anomalous dispersion in the telecom band, which has been a major challenge for $Si_3N_4$ films due to stress-induced cracking [25]. Recently, laser-reduced GO films with negative values of $n_2$ have been reported [55, 56]. In future work, it is anticipated that the use of rGO with a negative $n_2$ can reduce the phase mismatch in $Si_3N_4$ waveguides with normal dispersion, which would lower the requirements for achieving phase matching in normal-dispersion devices, thus rendering them capable of playing more important roles in nonlinear optical applications.

Slot waveguides, with enhanced light-matter interaction enabled by the strong light confinement in the subwavelength slot regions, provide a better structure to exploit the material properties of GO [28, 219]. Although GO has shown advantages in conformal coating integrated wire waveguides [63], this is still challenging for narrow slot regions with widths < 100 nm and heights > 200 nm. This is mainly limited by the size of GO flakes used for self-assembly, which is typically ~50 μm. By modifying the GO synthesis methods and using more vigorous ultra-sonication, GO flakes with smaller sizes can be obtained, which are expected to address this issue and enable the implementation of GO hybrid slot waveguides with significantly improved nonlinear optical performance. In addition to MRRs [54], other resonant device structures can be employed to enhance the light-GO interaction based on the resonant enhancement effect, such as subwavelength gratings [220], photonic crystal cavities [195], and whisper-gallery-mode cavities [221, 222].

Although there has been a lot of work investigating the nonlinear optical performance of GO and rGO, much of this has been semi-empirical. More physical insights, such as the anisotropy of the optical nonlinearity, the dependence of the nonlinear optical properties on the reduction / doping degree, and the interplay between Re ($\chi^{(3)}$) and Im ($\chi^{(3)}$) processes, remain to be explored. Previously, the optical nonlinearity of thick GO films (> 1 μm) was characterized via the widely used Z-scan method [55, 56], However, for extremely thin 2D films (< 20 nm), it is very difficult to accurately distinguish the weak response induced by the 2D films from the backgroud noise in the Z-scan measurements. Moreover, the ultrathin 2D films are easily damaged by the perpendicularly focused laser beam. The fabrication techniques for integrating

GO films allow for precise control of their thicknesses and sizes, which yields new possibilities for investigating fundamental physical insights of 2D GO films. This, in turn, will also facilitate the full exploitation of the great potential of GO in nonlinear integrated photonic devices. This synergy will have a long-lasting positive impact, which will be a strong driving force for the continuous improvement of device performance and broadening of applications.

Accompanying the continuous improvement in the knowledge and control of GO's material properties as well as the development of its fabrication techniques, it is expected that many new breakthroughs in GO nonlinear integrated photonics will happen, and indeed this has been a continuing very active field. [223-229] The delivery of mass-producible hybrid nonlinear integrated photonic devices with significantly improved performance serves the common interest of many photonic industries, which will accelerate the applications of 2D materials out of laboratory and assure that the research in this area will benefit the broader community. This material and technology will have a broad range of applications including integrated nonlinear photonic chips,[230-273] integrated microcombs,[274-302] quantum optics, [303-311] fibre-optic optical communications,[312-314] neuromorphic computing,[315-320] microwave photonics,[320-405] and many other areas.

## 7. Conclusion

The on-chip integration of GO films that provide a large optical nonlinearity along with a high degree of flexibility in changing its properties, represents a promising frontier for implementing high-performance nonlinear integrated photonic devices for a wide range of applications. In this paper, we review the progress in GO nonlinear integrated photonics. We summarize the optical properties of GO and the fabrication technologies for its on-chip integration. We review a wide range of GO hybrid integrated devices for different nonlinear optical applications, and compare the nonlinear optical performance of different integrated platforms. We also discuss the challenges and perspectives of this nascent field. Accompanying the advances in this interdisciplinary field, we believe that GO based nonlinear integrated photonics will become a new paradigm for both scientific research and industrial applications in exploiting the enormous opportunities arising from the merging of integrated devices and 2D materials.

**Conflict of Interest**

The authors declare no conflict of interest.